%% file: paper.tex
\documentclass[acmtog]{acmart}
\usepackage{colortbl}
\usepackage{xcolor}
\usepackage{float}
\usepackage{subcaption}
\usepackage{tabularray}
\usepackage{booktabs}

\input{macros}

\input{macro} % macros Beibei

\makeatletter
\newcommand\paragraphNew{\@startsection{paragraph}{4}{\parindent}%
  {-.5\baselineskip \@plus -2\p@ \@minus -.2\p@}%
  {-3.5\p@}%
  {\ACM@NRadjust{\@parfont}}}
	\makeatother
	
\AtBeginDocument{%
  \providecommand\BibTeX{{%
    \normalfont B\kern-0.5em{\scshape i\kern-0.25em b}\kern-0.8em\TeX}}}

\newcommand{\added}[1]{\textcolor{black}{#1}}
\newcommand{\remind}[1]{\textcolor{black}{#1}}

\newcommand{\beibei}[1]{\textcolor{black}{#1}}

\newcommand{\milos}[1]{\textcolor{black}{#1}}

\usepackage{amsmath}
\begin{document}

%%
%% The "title" command has an optional parameter,
%% allowing the author to define a "short title" to be used in page headers.

\title{Fiber-level Woven Fabric Capture from a Single Photo}

\author{Zixuan Li}
\orcid{0009-0004-2424-9529}
\authornote{Contribute equally.}
\affiliation{
    \institution{Nankai University}
    \city{Tianjin}
    \country{China}
}
\email{zixuan.li_2001@outlook.com}

\author{Pengfei Shen}
\orcid{0000-0002-2945-0115}
\authornotemark[1]
\affiliation{
    \institution{University of Science and Technology of China}
    \city{Hefei}
    \country{China}
}
\email{jerry_shen@mail.ustc.edu.cn}

\author{Hanxiao Sun}
\orcid{0009-0009-8887-2148}
\authornotemark[1]
\affiliation{
    \institution{Nankai University}
    \city{Tianjin}
    \country{China}
}
\email{hx.sun@mail.nankai.edu.cn}

\author{Zibo Zhang}
\orcid{0000-0002-1242-5183}
\affiliation{
    \institution{Nankai University}
    \city{Tianjin}
    \country{China}
}
\email{zibozhang@mail.nankai.edu.cn}

\author{Yu Guo}
\affiliation{
    \institution{George Mason University}
    \country{USA}
}
\email{tflsguoyu@gmail.com}

\author{Ligang Liu}
\affiliation{
    \institution{University of Science and Technology of China}
    \city{Hefei}
    \country{China}
}
\email{lgliu@ustc.edu.cn}

\author{Ling-Qi Yan}
\affiliation{
    \institution{University of California, Santa Barbara}
    \country{USA}
}
\email{lingqi@cs.ucsb.edu}

\author{Steve Marschner}
\affiliation{
    \institution{Cornell University}
    \country{USA}
}
\email{srm@cs.cornell.edu}

\author{Milo\v{s} Ha\v{s}an}
\orcid{0000-0003-3808-6092}
\affiliation{
    \institution{Adobe Research}
    \city{San Jose}
    \country{USA}
}
\email{milos.hasan@gmail.com}

\author{Beibei Wang}
\orcid{0000-0001-8943-8364}
\authornote{Corresponding author.}
\affiliation{
    \institution{Nanjing University}
    % \institution{Nanjing University of Science and Technology, Nankai University}
    \city{Suzhou}
    \country{China}
}
\email{beibei.wang@nankai.edu.cn}
% \renewcommand{\shortauthors}{Jin et al.}

%authornote and Nankai University
%%
%% The "author" command and its associated commands are used to define
%% the authors and their affiliations.
%% Of note is the shared affiliation of the first two authors, and the
%% "authornote" and "authornotemark" commands
%% used to denote shared contribution to the research.

%\author{Beibei Wang}
%\affiliation{School of Computer Science and Engineering, Nanjing University of Science and Technology}
%\author{Ling-Qi Yan}
%\affiliation{University of California, Santa Barbara}

%%
%% By default, the full list of authors will be used in the page
%% headers. Often, this list is too long, and will overlap
%% other information printed in the page headers. This command allows
%% the author to define a more concise list
%% of authors' names for this purpose.
%\renewcommand{\shortauthors}{Wang, et al.}

\begin{abstract}
%Capturing the appearance of woven fabrics is challenging, due to their complex detailed 3D microstructure that does not directly fit common material models, primarily designed to approximate reflection from surface heightfields. 
%Capturing the appearance of woven fabrics is challenging, because their complex 3D microstructure does not fit common material models, which are primarily designed to approximate reflection from heightfield surfaces. 
%Previous work \cite{Jin:2022:inverse} demonstrated a lightweight setup capable of recovering yarn-level woven fabric parameters, which achieves plausible appearances from distant views and allows for fast analytic rendering, but is less faithful for close-up views.

\milos{Accurately rendering the appearance of fabrics is challenging, due to their complex 3D microstructures and specialized optical properties. If we model the geometry and optics of fabrics down to the fiber level, we can achieve unprecedented rendering realism, but this raises the difficulty of authoring or capturing the fiber-level assets. Existing approaches can obtain fiber-level geometry with special devices (e.g., CT) or complex hand-designed procedural pipelines (manually tweaking a set of parameters). In this paper, we propose a unified framework to capture fiber-level geometry and appearance of woven fabrics using a \emph{single} low-cost microscope image. This may seem like an impossible task: a single microscope photo looks very different from the final rendering we would like to achieve, and the information contained in it may seem minimal. We propose a novel fiber parameter estimation pipeline in a coarse-to-fine manner, establishing a subset of parameters step by step. At the core of our pipeline are differentiable procedural geometric and appearance models for woven fabrics at the fiber level, enabling both geometry and appearance to be optimized simultaneously. We first use a simple neural network to predict initial parameters of our geometric and appearance models. From this starting point, we further optimize the parameters of procedural fiber geometry and an approximated shading model via differentiable rasterization to match the microscope photo more accurately. Finally, we refine the fiber appearance parameters via differentiable path tracing, converging to accurate fiber optical parameters, which are suitable for physically-based light simulations to produce high-quality rendered results. We believe that our method is the first to utilize differentiable rendering at the microscopic level, supporting physically-based scattering from explicit fiber assemblies. Our fabric parameter estimation achieves high-quality re-rendering of measured woven fabric samples in both distant and close-up views. These results can further be used for efficient rendering or converted to downstream representations. We also propose a patch-space fiber geometry procedural generation and a two-scale path tracing framework for efficient rendering of fabric scenes.}   
\end{abstract}

%%%Following previous work, we use a simple fabric capture configuration, wrapping the fabric sample on a cylinder of a known radius, capturing a single image under the known camera and light positions ad estimating coarse yarn parameters. 
%Starting from the coarse parameters, we propose to generate fibers procedurally from the center line of yarns and optimize the parameters of fiber geometric and shading models simultaneously to match the captured image better. The key insight is that, although the fibers are not clearly visible in the captured image, they must follow a specific distribution to produce the overall appearance. Therefore, valid fibers can be generated without precise visible fiber supervision. 
%%
%% The code below is generated by the tool at http://dl.acm.org/ccs.cfm.
%% Please copy and paste the code instead of the example below.
%%
\begin{CCSXML}
<ccs2012>
	 <concept>
	   <concept_id>10010147.10010371.10010372</concept_id>
		<concept_desc>Computing methodologies~Rendering</concept_desc>
		<concept_significance>500</concept_significance>
	 </concept>
   <concept>
       <concept_id>10010147.10010371.10010372.10010376</concept_id>
       <concept_desc>Computing methodologies~Reflectance modeling</concept_desc>
       <concept_significance>500</concept_significance>
       </concept>
 </ccs2012>
\end{CCSXML}

\ccsdesc[500]{Computing methodologies~Rendering}
\ccsdesc[500]{Computing methodologies~Reflectance modeling}
%%
%% Keywords. The author(s) should pick words that accurately describe
%% the work being presented. Separate the keywords with commas.
\keywords{fabric capture, fabric rendering, fiber-level}

%% A "teaser" image appears between the author and affiliation
%% information and the body of the document, and typically spans the
%% page.
\begin{teaserfigure}
\centering
\includegraphics[width=\textwidth]{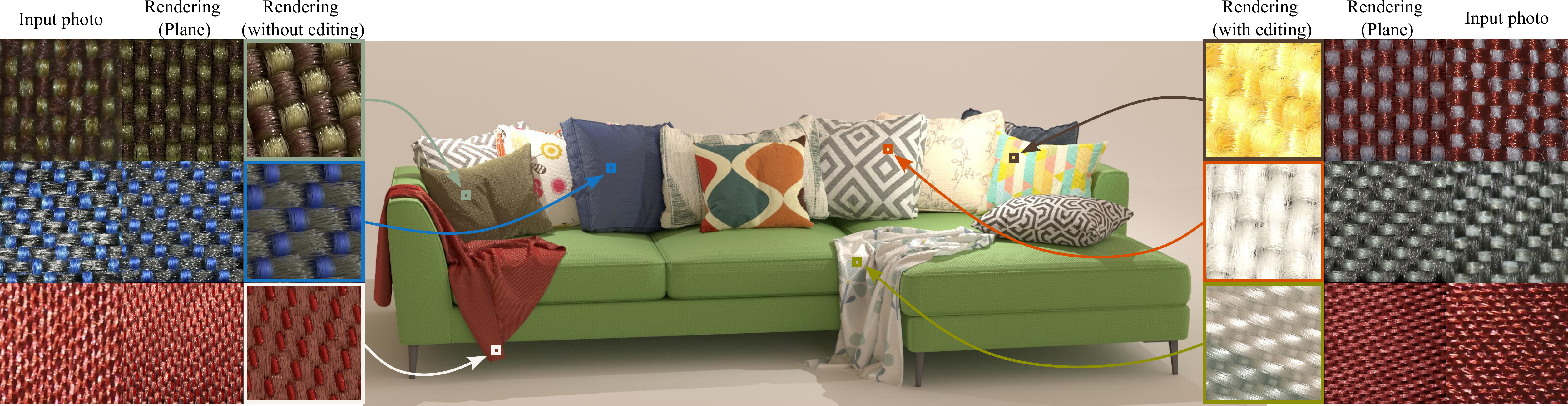}
\caption{Given a microscope photo of a woven fabric sample as input, our approach estimates the geometric and optical parameters of fibers in three coarse-to-fine steps: an initial geometry-appearance estimation via neural network, a joint geometry-appearance optimization with differentiable rasterization and an appearance refinement optimization with differentiable path tracing. Re-rendered results with the estimated parameters closely match the input photos (left-most and right-most columns). The resulting fabric parameters can be used for rendering directly or can be further edited to control the final appearance, as shown in the center. }
% \milos{MH: The editing is fairly drastic here, could we edit some examples slightly less? Also, the sizes of the squares in the image do not correspond to the insets directly, right?}} 
%The joint optimization of fiber geometric and appearance parameters with differentiable rasterization are provided
\label{fig:teaser}
\end{teaserfigure}
 %\mycfigure{pillow}{pillow.pdf}{}

%%
%% This command processes the author and affiliation and title
%% information and builds the first part of the formatted document.
\maketitle

\input{sec1_intro}
\input{sec2_related}
\input{sec3_background}

\input{table_parameters}

\input{sec4_forward}
\input{sec5_inverse}
\input{sec6_path}
\input{sec7_impl}

\input{sec8_results}
\input{sec9_conclusion}

%%
%% The acknowledgments section is defined using the "acks" environment
%% (and NOT an unnumbered section). This ensures the proper
%% identification of the section in the article metadata, and the
%% consistent spelling of the heading.
% \begin{acks}
% To Robert, for the bagels and explaining CMYK and color spaces.
% \end{acks}

%%
%% The next two lines define the bibliography style to be used, and
%% the bibliography file.
%\newpage 
\bibliographystyle{ACM-Reference-Format}
\bibliography{paper}
% Appendix
\appendix
\input{paper_supplementary}

\end{document}

%% file: macros.tex
\newcommand{\dd}{\mathrm{d}}

\newcommand{\bL}{\mathcal{L}}

\newcommand{\win}{\omega_\mathrm{i}}
\newcommand{\wout}{\omega_\mathrm{o}}

%% file: macro.tex
\def\figurePath{fig/}
\def\myfigure#1#2#3{\begin{figure}[tb]\centering\includegraphics[width = \linewidth]{\figurePath#2}\caption{#3}\label{fig:#1}\end{figure}}

\def\mycfigure#1#2#3{\begin{figure*}[tb]\centering\includegraphics*[clip, width = \linewidth]{\figurePath#2}\caption{#3}\label{fig:#1}\end{figure*}}

%% file: sec1_intro.tex
%-----------------------------------%
\section{Introduction}
\label{sec:intro}
%-----------------------------------%

Fabrics are essential to rendering applications ranging from digital humans to interior design. Recovering fabric appearance from photographs or other measurements has been studied for a long time, but remains challenging, due to the complex microstructure of fabrics. Repeated weave or knit patterns consisting of yarns, which themselves consist of highly specular fibers, are challenging and expensive to model at a highly detailed level. Moreover, perfectly repeating yarn patterns are unrealistic and hand-tuning the right amount of imperfection to achieve realism is tedious.

Existing approaches capture fabrics at different levels. Some make meso-scale (or yarn-level) approximations, while others go all the way down to  micro-level fiber representations. Fiber-level fabric models are capable of very high fidelity in renderings, but they generally require complex devices or pipelines to author~\cite{Zhao:2011:fabric, Khungurn:2015:matching, Zhao:2016:Yarn, schroder2015imagebased}. 
%For example, some methods capture the appearance through data-driven approaches such as bidirectional texture functions (BTFs)~\cite{kautz2007interactive}; other methods use micro-CT scanning approaches to capture the microstructures of the fabrics at the fiber level, which can produce highly detailed renderings~\cite{Zhao:2011:fabric, Khungurn:2015:matching, Zhao:2016:Yarn}, but a micro-CT scanner is not widely available. Some methods (\cite{schroder2015imagebased}) can recover the fabrics from a single image, but are based on a complex hand-designed pipeline and do not provide the optical properties of the fibers (only their geometry).
On the other hand, capturing yarn-level fabric parameters~\cite{Jin:2022:inverse, Tang:2024:pair} does not require expensive hardware, and fitting can be achieved cleanly by differentiable rendering. The recovered fabrics are, however, less faithful to close-up views without fiber details, as the forward model is not powerful enough to model this level of detail. 
Unlike these prior works, our work aims to fit fiber-level geometry and appearance from a single image captured with a widely available device in a unified way without manual tuning.
%This leads to the question: could a more powerful, fiber-level model still support differentiable rendering and recover woven fabrics at a much finer level of detail? % Another option is applying generic shading models not meant for fabrics, as single-image capture methods exist for such models~\cite{deschaintre2018single,Guo2021HighlightawareTN}. However, these produce less realistic results than using a fabric-focused model.

In this paper, we propose a fiber-level procedural fabric parameter estimation approach from a single image captured with a low-cost microscope camera. There is clearly no hope of exactly reproducing the observed fibers from such minimal input. The key insight of our solution is that given the supervision captured with a microscope camera, plausible fiber statistics can be procedurally generated in terms of geometries and appearance, even though they do not have to match the explicit fibers in the photo directly. In other words, the novelty of our approach is in capturing appearance by observing and fitting microscopic structure, as opposed to the traditional approach of measuring or fitting reflectance (BSDF).

The challenge is that the fiber geometries and optical parameters form an extensive parameter space. Meanwhile, the forward light transport involves long multiple scattering paths. Hence, the essential problem lies in how to estimate these parameters feasibly. To address these challenges, we introduce a novel fiber-level fabric parameter estimation pipeline in a coarse-to-fine manner.

Specifically, given a microscope image captured with known camera and light positions, we first use a neural network to predict the geometric and shading parameters. We then further optimize the geometry and appearance parameters jointly via differentiable rasterization, back-propagating the image differences using our geometric and shading models towards the model parameters. Finally, the estimated fiber optical parameters are further refined with differentiable path tracing, which would be too expensive and incapable of handling geometry optimization on its own.
Additionally, we provide an efficient fiber path tracing implementation, allowing efficient training data generation and large-scale scene rendering with our estimated fabric parameters.

To our knowledge, our work is the first to propose matching microscope photos using differentiable rendering with micro-scale geometric and appearance models. We capture highly detailed fabrics, showing realistic appearances from both distant and close-up views.

%% file: sec2_related.tex
\section{Related Work}
\label{sec:related}
%-----------------------------------%

In this section, we review fabric forward rendering models, fabric capture methods and procedural material parameter estimation.

\subsection{Fabric forward models}

%Existing approaches model the fabric appearance at different levels, from the yarn level to 

\paragraph{Surface fabric appearance models.}
Several previous approaches model fabric appearance using surface bidirectional scattering distribution function (BSDF) models (e.g., \cite{Adabala:2003:cloth, IrawanAndMarschner2012, Sadeghi:2013:Cloth, Deschaintre:23:visual_fabric}), producing a variety of fabric appearances. The SpongeCake surface appearance model by \citet{Wang:2021:Sponge} defines each layer as a volumetric medium described by a microflake distribution \cite{Jakob:2010:microflake, heitz2015SGGX} and can also be applied to fabrics, where the microflakes are typically oriented in alignment with fibers. Jin et al.~\shortcite{Jin:2022:inverse} use an extended version of the SpongeCake model, with further additions to enhance fabric realism. Zhu et al.~\shortcite{zhu:2023:cloth} improve this further, including rough and delta transmission, as well as shadowing-masking effects among yarns. Recently, \citet{Zhu:2024:multiscaleFabric} further extend this work to support multi-scale rendering and add more realistic sheen and parallax offset.

These types of BSDF models are very convenient in rendering systems but produce less realistic results for close-up views, due to missing geometric detail. Our model is built at the fiber level to produce more plausible results in close-up views.

\paragraph{Fiber scattering models.}
Another approach is to work at the fiber level and model the scattering function for each fiber with the bidirectional curve scattering distribution function (BCSDF) \cite{KajiyaAndKay:1989:hair,Marschner:2003:HairBCSDF,d'Eon:2011:hair}. We use the model of \citet{Chiang:2015:fur}, which instead gives the BSDF of any point on the fiber, not averaging over thickness. Computing multiple scattering among fibers by path tracing is expensive. Dual Scattering~\cite{Zinke:2008:dual} has been proposed to estimate multiple scattering among fibers; this solution is specifically designed for human hair but becomes less accurate for arbitrary fabrics. Recently, Zhu et al.~\shortcite{Zhu:2023:yarn} proposed a practical yarn-based shading model for cloth, by aggregating the light simulation on the fiber level. They also make several modifications (e.g., an additional diffuse component) to Dual Scattering to match fabrics. 

%In our paper, we model fabrics at the fiber level and estimate the appearance parameters of fibers. For that, we choose an approximated appearance model -- a single scattering hair BSCDF together with a diffuse term, to enable differentiability and efficiency for geometry and appearance estimation. \added{Then, we use optimization to obtain the standard BCSDF, ensuring high-fidelity renderings}.  

\subsection{Fabric capture methods}

We only review the most relevant work; please refer to the survey by Castillo et al.~\shortcite{Castillo2019recent} for more work in this area.
We specifically focus on woven fabric recovery, leaving knitted fabric recovery~\cite{trunz2019inverse, kaspar2019neural} for future work. Our model does not consider color pattern variations; these can be added by further parameter editing; we could add them by combination with previous work~\cite{rodriguez2019automatic}.

%\added{From Steve: Seems like you want surface, yarn, and fiber level in the forward section, and then you can use the same structure again in discussing capture methods.}

\paragraph{Surface-level fabric capture models.}

From a single image captured with a cell phone camera, Jin et al.~\shortcite{Jin:2022:inverse} estimate initial fabric parameters and further optimize using differentiable rendering to refine the recovered parameters. Their work can match the captured image well from a distant view, but shows less plausible rendering at close-up views, due to lack of fiber-level details. Recently, Tang et al.~\shortcite{Tang:2024:pair} improve previous forward surface shading model and recover both the reflection and transmission appearance of the fabric. They capture a pair of reflection and transmission images, recovering fabric parameters using a pipeline similar to the method of Jin et al.~\shortcite{Jin:2022:inverse}. Their work matches both reflection and transmission, but similarly to Jin et al. lacks fiber details in close-up views. It may be interesting to combine our approach with Tang et al.'s to match a pair of reflection-transmission photos at the fiber level.

Garces et al.~\shortcite{Garces:2023:Dual-scaleOpticalSystem} capture both microscopic and macroscopic images, and match simultaneously at both scales. They use a modified anisotropic Disney model \cite{Burley:15:Disney} and show accurate fits at the micro-scale as well as meso-scale. However, they use a custom hemispherical device incorporating multiple cameras of different scales and various types of lighting, rather than a cheap off-the-shelf device like in our case.

\paragraph{Fiber/ply-level fabric capture models.}
In contrast to surface-level capture models, another group of approaches capture fiber-level or ply-level geometries. These approaches can further be categorized into two types, depending on whether they rely on specialized devices or not. Several methods~\cite{Zhao:2011:fabric, Khungurn:2015:matching, Zhao:2016:Yarn} use micro-CT scanning approaches to capture the microstructures of fabrics at the fiber level, producing highly detailed renderings; however, the capture is expensive. Similar to our work, Khungurn et al.~\shortcite{Khungurn:2015:matching} use differentiable rendering, but only for appearance model parameters; \added{the geometry still still comes from micro-CT scans and is not differentiable}. Zhao et al.~\shortcite{Zhao:2016:Yarn} propose a procedural model to generate fibers for a yarn, considering the cross-sectional fiber distributions and migrations. A recent work by Montazeri et al.~\shortcite{Montazeri:2020:ply} models the detailed geometries at the ply level, resulting in a practical yet accurate fabric geometry recovery model. They also introduced an aggregated fiber scattering model, allowing reflection and transmission. However, their appearance cannot be captured automatically. Similar to previous work \cite{Zhao:2016:Yarn, Montazeri:2020:ply}, our fiber geometries are generated procedurally, but with some simplifications, to enable efficient differentiable rendering.

%\paragraph{Single-image fabric recovery.}
Unlike the methods that rely on specialized devices, several approaches can reverse-engineer cloth from standard photos. Schr{\"o}der et al.~\shortcite{schroder2015imagebased} estimate yarn geometry from a single image; with manual selection of model parameters, they achieve fiber-level details and produce visually plausible results. Guarnera et al.~\shortcite{Guarnera2017wfmc} recover high-quality yarn-level parameters via the spatial and frequency domain, but without any fiber details. %Then, they estimate the optical parameters with an image captured from a cell phone. There are several differences between their approach and ours. First, their approach estimates the geometry and the appearance with different input images, while our approach only relies on a single microscope image and estimates both the geometry and the appearance in a unique automatic pipeline, resulting in a much simpler solution. Second, their appearance model is a surface-based model, which is less photorealistic compared to the fiber scattering model used in our method. 

\added{A closely related work to ours is by Wu et al.~\shortcite{Wu2019modeling}, which estimates yarn-level geometries from a single micro-image captured with a microscope camera and enhances the fine-scale fiber details with the fiber twist. However, their appearance parameters are manually tweaked. In contrast, our work recovers both fiber geometries and appearance via differentiable rendering.}

\subsection{Procedural material parameter estimation.}

Several works have been proposed for procedural material parameter estimation or generation by learning a mapping from images to material parameters with neural networks~\cite{hu2019novel}, a Bayesian framework~\cite{Guo:2020:Bayesian}, or a differentiable version of Adobe Substance material graphs~\cite{Shi:2020:MATch}. Recent work~\cite{Guerrero:2023:MatFormer} proposes to use a transformer network to generate procedural material graphs; this was later extended to conditioned generation with text prompts or images~\cite{Hu:2023:Procedural}.

The above methods are designed for general materials and are not specialized nor optimal for fabrics, nor for the level of detail we target. Since our method focuses on reconstructing woven fabrics, we use procedural fabric geometry and a shading model appropriate for woven fabrics. We also propose a different capture setup, using microscope images with known lighting to capture the fine details of fabrics.

%% file: sec3_background.tex
\section{Background and overview}
\label{sec:Background}
%-----------------------------------%

Woven fabrics are manufactured by a process that weaves weft yarns through stretched parallel warp yarns in a specific repeating pattern. Each yarn is made of fibers, and we need a realistic model to shade a single fiber.

In this section, we review the core fiber appearance model used in our framework~\cite{Chiang:2015:fur} and then formulate our problem.

\subsection{Fiber single scattering model}

The single fiber scattering model of Chiang et al. \shortcite{Chiang:2015:fur} separates scattered light into multiple modes based on the number of internal light reflections within the fiber:
\begin{equation}
    S(\theta_\mathrm{i},\theta_\mathrm{o},\phi)=\sum_{p=0}^{\infty}S_p(\theta_\mathrm{i},\theta_\mathrm{o},\phi).
\end{equation}
The $S_p(\theta_\mathrm{i},\theta_\mathrm{o},\phi)$ denotes the $p$-th scattering lobe, which accounts for all light paths that undergo $p-1$ internal reflections inside the fiber cylinder ($\{R=0, TT=1, TRT=2, TRRT=3, \cdots\}$).
Each lobe is represented as a product of the longitudinal scattering function $M_p$ and the azimuthal scattering function $N_p$:
\begin{equation}
    S_p(\theta_\mathrm{i},\theta_\mathrm{o},\phi)=M_p(\theta_\mathrm{i},\theta_\mathrm{o})N_p(\phi).
\end{equation}
Unlike previous fiber shading models \cite{Marschner:2003:HairBCSDF, d'Eon:2011:hair}, the azimuthal scattering function is no longer integrated over the fiber width. Instead, the offset across the fiber $h$ is used as a parameter, and the integral over the fiber width is directly computed by pixel filtering in the rendering system:
\begin{equation}
    N_p(\phi,h)=A_p(h)D_p(\phi-\Phi(p,h)).
\end{equation}
Thus this model describes a traditional BSDF for a shading point on a fiber modeled as an explicit surface, rather than a BCSDF assuming infinitely thin fibers, like in some previous models. For more details, see the paper \citet{Chiang:2015:fur}.

\subsection{Pipeline Overview}

\label{subsec:problem_formulation}

\myfigure{config}{configration.pdf}{Our capture configuration for real fabric data. We use a microscope camera to capture the fabric samples under one of the light sources inside the microscope, blocking off other lights. The fabric sample is placed on a plane under the camera. We measure the distances between all elements, as well as the {camera field of view}, so we can reconstruct the same setup in synthetic renderings. }
% \milos{MH: The color of the fabric does not match between left and center vs. right.}}
%The configuration to measure the real fabric data. We use a cell phone flash as a light source, though any point light source can be used. The fabric sample is wrapped onto a cylinder. The distances between elements are measured, as well as the cylinder diameter and camera field of view, so we can reconstruct the same setup in synthetic renderings.

Fiber-level models can obviously achieve more realistic renderings than surface-based models. However, the fibers of woven fabrics are tiny enough that they are not explicitly visible in normal photos from commonly used cameras. Therefore, we opt for a low-cost microscope camera as our capture device. Our work aims to reconstruct fiber geometries with perceptually matching appearance, given a photo captured under a known microscope camera with a known light source, as shown in Fig.~\ref{fig:config}. 

As we aim to reconstruct both fiber geometries and materials, we need to define the geometric and appearance model. Defining each fiber curve explicitly would require an immense parameter space. For example, a small fabric sample patch (1 cm $\times$ 1 cm) consists of about 8,000 fibers and 800,000 control vertices. A procedural geometric model can reduce the parameter space significantly: each yarn can be coarsely represented with a centerline curve and a set of fibers around the centerline can be procedurally generated based on a small number of parameters (see Fig.~\ref{fig:yarnfiber} and Table~\ref{tab:param}). We currently assume all weft fibers are optically identical, and the same for all warp fibers. %Next, we also need to establish the shading model of fibers. As we target high-fidelity renderings of fibers, a fiber single scattering model, i.e., BCSDF, is required for downstream applications. 
%The main question here is how to compute multiple scattering. The most accurate way requires expensive path tracing with a long path, making it infeasible to reconstruct the geometry and the appearance together. Therefore, a simpler, computationally efficient, and differentiable shading model is needed for the geometry and coarse appearance reconstruction. 

After defining the above fiber parameter set, the problem of constructing fiber geometries and optical properties that match our target images can be formulated as an optimization problem, minimizing the difference between the rendered image $I$ and the target image $T$:
\begin{equation}
   \mathop{\min}_{\mathcal{S}} E_{\text{loss}}\left(I(\mathcal{S};V,L),T \right),
     \label{eq:opt}
\end{equation}
where $T$ is the target microscope photo, $I(\mathcal{S};V,L)$ is the predicted (rendered) result with parameters $\mathcal{S}$ under view $V$ and light $L$, and $E_{\text{loss}}$ is a loss function to measure the distance between rendered and target images. We utilize a perceptual loss to evaluate image similarity without requiring pixel alignment, a color loss to reduce color bias, and a prior term to improve optimization stability.

We apply differentiable rendering to this optimization problem, handling both fiber geometry and appearance in a unified way; this is a key difference from previous work, which requires manual appearance matching~\cite{schroder2015imagebased} or has a simple appearance matching but a complex geometry matching step~\cite{Khungurn:2015:matching}. Our fiber parameter estimation method is presented in Sec.~\ref{sec:inverse}.

Finally, we also need a way to efficiently render scenes that consist of fiber-level fabrics at real-world scales. The final forward renderings require massively more fibers and control vertices than the microscopic sample used for inverse rendering. For this, we introduce a practical two-scale path tracing framework (Sec.~\ref{sec:path}).

%Note that the light and the view directions are known, and the fiber parameters $\mathcal{S}$ will be optimized. 

%The key insight of our solution is that given the supervision captured with a microscope camera, plausible fiber statistics can be procedurally generated in terms of geometries and appearance, even though they do not have to match the explicit fibers in the photo directly.

%\paragraph{Formulation}

%Woven fabrics are manufactured by a process that weaves weft yarns through stretched parallel warp yarns in a specific repeating pattern. In our method, a fabric sample is represented as a set of yarns. 

%

%\paragraph{Challenges.}
%Our goal is to optimize the fiber parameters $\mathcal{S}$ to minimize Eqn.~(\ref{eq:opt}). 
%Previous work does not estimate parameters within a unified optimization framework, and requires manual appearance matching~\cite{schroder2015imagebased} or has a simple appearance matching but a complex geometry matching step~\cite{Khungurn:2015:matching}. To overcome these issues, we introduce differentiable rendering into this optimization problem, handling both fiber geometry and appearance in a unified way. Since the fiber geometric and appearance models are designed for optimization, and large numbers of fibers are needed even for small fabric sample patches (e.g., a 1 cm $\times$ 1cm fabric consists of about 2,000 fibers and 200,000 vertices), both the geometric and appearance models must be physically realistic, differentiable, computationally efficient, and described by a compact set of parameters.

%% file: table_parameters.tex
\begin{table}[!t]
	\renewcommand{\arraystretch}{1.5}
	\caption{\label{tab:param} 
Parameters in our model. Warp yarns are denoted with $v$-subscripts (vertical) and weft yarns with $h$-subscripts (horizontal). The top twelve parameters affect yarn geometry, and the rest affect appearance. Here, ``neural'' indicates predicted with the neural network, ``DR'' means predicted with the differentiable rasterization, and ``DPT'' means predicted with the differentiable path tracing.}
\begin{large}
 % \left
  \scalebox{0.81}{
  \begin{tabular}{|l|cc|l|}\hline
  %       $s^\mathrm{h},s^\mathrm{v}$ & yarn size for weft and warp yarns \\
		% $\beta^\mathrm{h}, \beta^\mathrm{v} $ & heightfield scaling factor for weft and warp yarns \\ \hline
		% $k_\mathrm{d}^\mathrm{h}, k_\mathrm{d}^\mathrm{v}$  &diffuse albedo for weft and warp yarns  \\
		% $k_\mathrm{s}^\mathrm{h}, k_\mathrm{s}^\mathrm{v}$  &specular albedo for weft and warp yarns  \\
		% $\alpha^\mathrm{h}, \alpha^\mathrm{v} $ & roughness for weft and warp yarns \\
		% $\psi $ & fiber twist angle \\
		% $u_{\mathrm{max}} $ & maximum inclination angle \\
		% % $T$ & thickness of the fabric \\
		% $w$ & weight for the Lambertian term blending \\
		% $U_\mathrm{s}(\xi)$ & randomness on the specular term \\
		% $U_\mathrm{n}(\xi)$ & randomness on the normal and orientation \\
		% $\mathrm{Q}$ & normal / orientation randomness level \\ \hline
        % $\beta,\epsilon$ & cross-sectional fiber distribution \\
        $T$ & fabric pattern & neural\\
        $L^\mathrm{h}, L^\mathrm{v}$ & fabric physical size & input\\
        %$\alpha, R_i, \theta_i$ & fiber twisting \\
        $n^\mathrm{h}, n^\mathrm{v}$ 
        & {yarn count per fabric}& neural \\
        $l^\mathrm{h}, l^\mathrm{v}$ & {yarn segment length} & - \\
         $u_{\mathrm{max}}^\mathrm{h}, u_{\mathrm{max}}^\mathrm{v} $ & {maximum inclination angle}  & neural + DR \\
        $\beta^\mathrm{h}, \beta^\mathrm{v}$ & heightfield scaling factor & neural + DR\\
        $e_\mathrm{yarn}^\mathrm{h}, e_\mathrm{yarn}^\mathrm{v}$ & yarn radius & neural\\
        $m^\mathrm{h}, m^\mathrm{v}$ & fiber count per yarn & neural\\
        ${G^\mathrm{h}, G^\mathrm{v}},s^\mathrm{h}, s^\mathrm{v}$ & fiber migration & neural\\
         $\alpha^\mathrm{h}, \alpha^\mathrm{v}$ & fiber twisting & neural + DR\\
         $Q^\mathrm{h},Q^\mathrm{v}$ & noise level & DR
        \\ \hline
         $C^\mathrm{h}, C^\mathrm{v}$ & fiber albedo coefficients & neural + DR + DPT\\
         $\gamma_\mathrm{M}^\mathrm{h}, \gamma_\mathrm{M}^\mathrm{v}$ & longitudinal roughness & neural + DR + DPT\\
         $\gamma_N^\mathrm{h}, \gamma_N^\mathrm{v}$ & azimuthal roughness & neural + DR + DPT\\
         $\gamma_\mathrm{M0}^\mathrm{h}, \gamma_\mathrm{M0}^\mathrm{v}$ & primary reflection roughness & neural + DR + DPT\\
         $k_\mathrm{d}^\mathrm{h}, k_\mathrm{d}^\mathrm{v}$  &diffuse albedo for fibers  & DR \\
         $w_\mathrm{d}$ &  diffuse term blending weight & DR\\
%\beibei{$F_\mathrm{TRT}^\mathrm{h},F_\mathrm{TRT}^\mathrm{v}$} & \beibei{TRT lobe factor} &  opt\\
        %\beibei{$d^\mathrm{h},d^\mathrm{v}$} & %\beibei{scattering density factor} &  opt\\
			\hline
		\end{tabular}}
\end{large}
\end{table}

%% file: sec4_forward.tex
%-----------------------------------%
\section{Differentiable fiber geometric model}
\label{sec:forward}
%-----------------------------------%

%\added{Beibei is working on this section and will finish in one day..}

%For that, we propose practical fiber geometric and shading models in our paper. 

%To this end, we propose a procedural geometric model for fibers, where the fibers are generated for each yarn with several parameters. Regarding the appearance model of fibers, one option is model it with the scattering function for individual fibers, or called hair BSDF. However, the multiple scattering in the fiber need path tracing, which leads to noisy and expensive time cost. The other option is using the multiple-scattering approximation, like dual scattering, which provides plausible quality, and the time cost is much less than the single-fiber scattering model. Therefore, our final choice is a procedural fiber generation model for geometry and the dual scattering model for appearance.

\myfigure{yarnfiber}{centerline.pdf}{(a) Woven fabric consists of weft and warp yarns. (b) The position of each yarn is coarsely defined by a centerline curve. The shape of a centerline curve is defined by several parameters: the maximum inclination angle $u_\mathrm{max}$, the length of a yarn segment, and the radius of the circle. (c) Around each centerline curve, a set of fibers is generated.}

% \added{We define the $y$ axis as the yarn direction and the $z$ axis as the normal to the rectangular fabric sample. Our notation is summarized in Table~\ref{tab:param}.}

% \subsection{Differentiable fiber geometric model}
% \label{sec:geometry}

In this section, we introduce our procedural geometric model for woven fabric fibers. There are several key requirements on the model, including matching real captures closely, being differentiable, and having a compact set of parameters. We propose a hybrid analytic formulation for the yarn-level centerline curve, together with a randomized fiber generation strategy. 

\subsection{Notations}
%We first define the coordinates of the yarn and fibers and then denote the symbols. 
A centerline curve of a yarn segment gives the height (displacement) $z$ as a function of position along the $x$ or $y$ axis for weft and warp, respectively. We will use $y$ below without loss of generality. As shown in Fig.~\ref{fig:yarnfiber} (c), fibers are twisted along the centerline curve with a twist angle $\alpha$, and the rotation angle $\theta$ varies with the $y$-coordinate along the curve as $\theta = \frac{2\pi y}{\alpha}$.

%We first define the centerline curve of each yarn and then generate fibers along these curves. 

%The centerline curves of previous works~\cite{Zhao:2016:Yarn,Montazeri:2020:ply} are undifferentiable. %Zhao et al. \shortcite{Zhao:2016:Yarn} generate the yarn curves by uniformly sampling in the UV space of a given base mesh. Montazeri et al. \shortcite{Montazeri:2020:ply} optimize the yarn curves using a physical model. However, their methods are all undifferentiable.
%Different from their work, we introduce a yarn centerline curve that can generate structures similar to real capture and is also differentiable. Then, we use the fiber generation model proposed by Zhao et al.~\shortcite{Zhao:2016:Yarn}, along with the cross-sectional fiber distribution function proposed by Trunz et al.~\shortcite{trunz2023neural}. Finally, we add new fiber-level noise to enhance randomness.

%\myfigure{parameters}{centerline_detail.pdf}{$l$ is defined as the length of the yarn segment, and $r$ is defined as the radius of the centerline arc. \added{Zibo: merge to figure 3, and include the $u_\mathrm{max}$ in., include the axis in, the y is horizontal, and z is the vertical axis. the center point is at the middle of the segment.}}

\subsection{Hybrid analytical yarn-level centerline curve}

Looking at the captured microscope images of several typical woven fabrics (rightmost, Fig.~\ref{fig:Ablation_geo}), we have several observations about the behavior of the yarn centerline curve: for the satin fabric, the long yarn segments in a satin weave pattern have large inclination angles combined with long flat center portions; for the plain pattern, yarns tend to be arc-shaped. Based on these observations, the proposed centerline curve should be able to cover all these typical height profiles.
%Our insight is that a circular arc can represent an arc-like shape at the center with a small inclination angle on the side, while a parabola-weighted circular arc (defined below) can express a long flat center with a large inclination angle on the side. Therefore, 
We propose a hybrid yarn-level centerline curve that combines these goals.
Specifically, we interpolate (blend) a circular function $z_\mathrm{cir}(y)$ with a parabolic function $z_\mathrm{par}(y)$, where the blend weight is based on the segment length. This hybrid formulation enhances the inclination at the endpoints while allowing a flatter center:
\begin{equation}
    \begin{aligned}
    z_\mathrm{cir}(y) &= \sqrt{r^2 - (y-r\mathrm{sin}(u_\mathrm{max}))^2} - r \mathrm{cos}(u_\mathrm{max}), \\
    z_\mathrm{hyb}(y) &= w z_\mathrm{cir}(y)z_\mathrm{par}(y) + (1-w)z_\mathrm{cir}(y),\\
    z_\mathrm{par}(y) &= \frac{\left(y-r\sin\left(u_\mathrm{max}\right)\right)^{2}}{rl}+1,\\
    w &=\frac{\left(l-l_\mathrm{min}\right)}{(l_{\mathrm{max}} {- l_{\mathrm{min}}})},
\end{aligned}
\end{equation}
where $u_\mathrm{max}$ is the inclination angle at the endpoints of the yarn segment. $r$ is the radius of the circle and can be computed as $r = \frac{l}{2\mathrm{sin}(u_\mathrm{max})}$.
$z_\mathrm{par}(y)$ is the parabola function defined in position $y$, $w$ is the weight of the interpolation, and $l$ is defined in $[l_\mathrm{min},l_\mathrm{max}]$, set as $l_\mathrm{min} = 1$ and $l_\mathrm{max} = 4$ in practice (where one length unit covers one cell of a weave pattern). In Fig.~\ref{fig:Ablation_geo}, we demonstrate the comparison between two curves and their blending with the microscope captures.

Finally, we perform a normalization by dividing out the maximum height of the function to decouple the effect of the yarn segment length and the height, and multiply with an additional height scaling factor $\beta$ to arrive at our yarn centerline formulation:
\begin{equation}
\label{eq:centerline}
    z_\mathrm{cen}(y) = \frac{\beta z_\mathrm{hyb}(y)}{r-r\cos\left(u_\mathrm{max}\right)}.
\end{equation}
We show several centerline curves in Fig.~\ref{fig:curves} for varying parameters. Our function can represent a wide range of yarn profiles by setting the three parameters $u_\mathrm{max}$, $l$ and $\beta$, showing a sufficient representation ability. 

\myfigure{Ablation_geo}{Ablation_geo.pdf}{Comparison of different curves. The last column shows the ground truth obtained by microscope capture \cite{Sadeghi:2013:Cloth}, with purple lines indicating the centerlines of the yarns, which are treated as the ground truth. For plain, parabola-weighted circular function causes the top of the yarn to be too flat. For satin, the circular function leads to unrealistic arching at the top of the yarn. The hybrid way avoids these issues.}

\myfigure{curves}{curve.pdf}{With different parameters ($u_\mathrm{max}$, $l$, and $\beta$), our centerline formulation Eqn.~(\ref{eq:centerline}) exhibits various shapes. The baseline settings are $u_\mathrm{max}=0.5\pi$, $l=2$ and $\beta = 1.0$.}

\subsection{Randomized fiber generation}

With a defined centerline curve, we need to further generate fibers around this centerline curve, e.g. to determine the position of warp fibers in the $x$-$z$ plane at a given $y$-coordinate (and similar for weft). Following previous work~\cite{Zhao:2016:Yarn}, a radius around each centerline curve at the cross-section plane needs to be established, together with some migration of this radius to introduce some randomness. However, their exact approach is unsuitable for our problem, as their generation operation is not differentiable, and the generated fibers still lack sufficient irregularity. Fibers also exhibit random behavior along the vertical and azimuthal directions due to the overlapping of yarns. To address these two issues, we introduce the cross-sectional fiber distribution function by Trunz et al. \shortcite{trunz2023neural} to enable differentiability, and introduce two types of noise to enhance randomness. 

Specifically, we define the cross-sectional fiber distribution function $R_i$ and introduce the migration (following Zhao et al.~\shortcite{Zhao:2016:Yarn}), by extending $R_i$ to depend on $\theta$:
\begin{equation}
 \begin{aligned}
  %  p(R) = (1-2\epsilon)\left(\frac{e-e^R}{e-1}\right)^\beta + \epsilon.
  R_i &= e_{\mathrm{yarn}}\left( \frac{i^{0.3}}{m^{0.3}} + J_{xy}S \right),\\
  R_{i}(y) &= (1 - G)R_{i} + \frac{GR_{i}}{2}\left[ \mathrm{cos}(s\theta+\theta_{i}^{(0)}) + 1\right],
 \label{eq:cross}
  \end{aligned}
\end{equation}
where $e_\mathrm{yarn}$ is the yarn radius, $m$ is the fiber count, $J_{xy}$ is a jitter amount for the sampled fiber by uniform sampling within $(0,0.1)$, and $S$ is a normal distributed random variable with zero mean and a standard deviation of 1. $G$ represents the extent of migration, defined in the range of $(0,1)$, and $s$ controls the length of a rotation. $\theta_i^{(0)}$ is a per-fiber parameter indicating the initial rotation, randomly distributed in the range $(0, 2\pi)$.

With the per-fiber distribution $R_{i}(y)$, a fiber curve can be generated by rotating along the centerline curve as a circular helix parameterized by $\theta$:
\begin{equation}
\begin{aligned}
    x_{\mathrm{fib}}(y) &= R_i(y) \mathrm{cos}(\theta+\theta_i) ,\\ z_{\mathrm{fib}}(y) &= R_i(y) \mathrm{sin}(\theta+\theta_i) + z_\mathrm{cen}(y),\\ 
    \theta_i &= 2 \pi \zeta i,
    \end{aligned}
\end{equation}
where $\zeta$ is a heuristically chosen constant to create a slightly pseudo-random distribution, and set as 0.137 in practice, following Trunz et al.~\shortcite{trunz2023neural}.

To enhance the irregular nature of the fibers, we further introduce two noise functions to the fibers, including a Perlin noise $z_\mathrm{P}$ along the vertical direction of each segment of the fibers, and a fiber-specific azimuthal noise on the centerline curve $z_\mathrm{cen}(y)$. These two types of noise can represent the random displacements of fibers along the vertical and azimuthal directions caused by the mutual tension between the weft and warp yarns, leading to the final formulation of $z_{\mathrm{fib}}(y)$:
\begin{equation}
\begin{aligned}
    z_{\mathrm{fib}}(y) &= R_i(y) \mathrm{sin}(\theta+\theta_i) + z'_\mathrm{cen}(y) + z_\mathrm{P},\\
    z_\mathrm{cen}'(y) &= \frac{\beta z_\mathrm{hyb}(y^{\kappa})}{r-r\cos\left(u_\mathrm{max}\right)},\\
    z_\mathrm{P} &= P({\lambda_\mathrm{ply}Q },{Q \lambda_\mathrm{fib}},{Q \lambda_\mathrm{seg}} + \xi), 
  %  \xi &= \mathrm{random}(0,100),\\
    \end{aligned}
\end{equation}
where $P$ is a three-dimensional Perlin noise function calculated on the weft and warp separately. $Q$ is the noise level in the range of (0,1), and $\lambda_x (x \in \{ ply,fiber,segment\})$ is the index of three types. Furthermore, we have incorporated an additional random noise offset $\xi$ to alleviate artifacts resulting from excessive continuity in the noise, where $\xi = \mathrm{random}(0,100)$. Regarding the azimuthal noise $z'_\mathrm{cen}(y)$, we randomly select the value of $\kappa$. When ${\kappa = \frac{5Q}{2l^2} + 1}$, the movement reaches its maximum. The impact of $\kappa$ on the fiber curve is shown in Fig.~\ref{fig:curves_movement}.

\myfigure{curves_movement}{curve_movement.pdf}{The effects of different $\kappa$ on the fiber curve. Our azimuthal noise causes the highest point of the fiber to move in the azimuthal direction. The farther $\kappa$ deviates from 1, the greater the movement. }

\mycfigure{completed_pipeline}{fullpipeline.pdf}{{Overview of our fiber parameter estimation.} Given an image captured with the microscope camera, we predict the fiber parameters (for a hair BCSDF model) with a prediction network. Then, we further optimize the parameters with differentiable geometric and shading models, where the shading model is the hair BCSDF for the single scattering and a diffuse term for the multiple scattering. Eventually, the approximated shading parameters (a single scattering and the diffuse term) are mapped into the single scattering parameters with an optimization with differentiable path tracing.}

%% file: sec5_inverse.tex
\section{Fiber parameter estimation}
\label{sec:inverse}
%---------------------------------------------%

%\revise{Note that performing a joint optimization of the geometry and the appearance in a differentiable path tracing framework is infeasible, since it requires an extremely long time and a huge storage space due to the large parameter space. Therefore, we optimize the geometric and appearance parameters with an approximated shading model in a differentiable rasterization pipeline. After getting the geometry parameters, optimizing the appearance parameters in a differentiable path-tracing framework becomes possible. }

\subsection{Fiber parameters}
\label{sec:fiber_param}

The parameters for fiber geometry include three parameters ($l$, $u_\mathrm{max}$ and $\beta$) for the centerline curve and six parameters (m, $e_\mathrm{yarn}$, $\alpha$, $s$, $G$, and $Q$) for randomized fiber generation. The parameters of the fiber appearance model are the longitudinal roughness $\gamma_M$, the azimuthal roughness $\gamma_N$, the primary reflection roughness $\gamma_{M_0}$ and the albedo $C$.

All parameters (except for the noise level $Q$) will be initialized by a neural network. Some parameters will be further refined by an optimization step, including all appearance parameters and some geometric parameters ($u_\mathrm{max}$, $\alpha$, $\beta$, and $Q$).

\subsection{Measurement setup for fabrics}
\label{sec:capture}

We propose a simple configuration to capture real fabric samples with a single microscope photograph, as shown in Fig.~\ref{fig:config}. We use one of the eight built-in light sources that come with the microscope. The fabric sample with physical size $L^\mathrm{h} \times L^\mathrm{v}$ is placed on a plane under the camera. The captured images have a resolution of $1280 \times 720$ and are cropped into a $720 \times 720$ square.

We measure the relative locations of the plane, camera, and light source to render images in the same synthetic setup. We use a point light in our rendered image to match the small LED light in the microscope camera. The field of view of the camera is also measured. We capture images with auto-exposure. We adjust the light brightness in our synthetic setup to match that of the captured image, initially setting the point light intensity to 100 and then optimizing it. 
%The details are described in Sec.~\ref{sec:captureDetails}.

%indicating that the pattern type and the yarn count will not be optimized.

%All these parameters are differentiable, and we optimize these parameters to match the input image and the rendered image of fibers rendered with these parameters.

%\added{Note that, we do not optimize the centerline curve of each yarn, since \added{xxx}.}

\subsection{Neural networks for parameter initialization}
\label{sec:network}

\myfigure{network}{network2.pdf}{Our network architecture. Given a target image, we estimate its fabric parameters with a neural network. More specifically, we extract the Gram matrix represented features from the target image with a pretrained VGG-19 and then perform a two-layer MLP to output the fiber parameters. Among the predicted parameters, we compute a cross-entropy loss for the fabric pattern and an $\bL_1$ loss for the other \remind{28} parameters.} 

Given a single image captured with the setup shown in Fig.~\ref{fig:config}, we estimate initial parameters with a simple neural network, which is similar in architecture to Jin et al.~\shortcite{Jin:2022:inverse}, but produces different parameters and uses different training data. 

%\paragraph{Architecture.} 
Our network is trained on synthetic data that includes several typical patterns. We first feed the input image into a pretrained VGG-19 network, which outputs the Gram matrices of several chosen layers. Later, the Gram matrices are fed into a fully connected (FC) module, which includes two intermediate layers (256 nodes per layer) with a LeakyReLU activation function. The final FC layer outputs the predicted parameters (29 channels for our geometric and appearance models). A sigmoid is used to output the parameters in the $(0,1)$ range, after which they are remapped into their respective ranges.

%\paragraph{Dataset generation.}
The loss function for network training consists of two components: a cross entropy loss between the ground truth fabric pattern index and the network predicted fabric pattern index (using one-hot encoding and softmax), and an $\bL_1$ loss for the other 28 continuous parameters. The details of the dataset and training are shown in Sec.~\ref{sec:netDetails}.

\paragraph{Discussion}
% \beibei{which parameters are only decided by the network? why? }
{The parameters of fabric pattern $T$, yarn count per fabric $n$, yarn radius $e_{\mathrm{yarn}}$, fiber count per yarn $m$, fiber migration $G$ and $s$ are only decided by the network. These parameters can be accurately predicted by the network, so we do not further optimize them to reduce the parameter space in the next optimization steps.}

Note that we do not predict the length of the yarn segment directly but instead estimate the yarn counts $n^\mathrm{h}, n^\mathrm{v}$. The length of the yarn segment can then be derived from the yarn count with $l^\mathrm{h} = \frac{L^\mathrm{h}}{n^\mathrm{v}}, l^v = \frac{L^\mathrm{v}}{n^\mathrm{h}}$, where $L^\mathrm{h}, L^\mathrm{v}$ is the fabric physical size.

\subsection{Joint geometry-appearance optimization via an approximated shading model}
\label{sec:DR}

\myfigure{opt_pipleline}{optimize_overview.pdf}{Overview of our joint geometry-appearance optimization. Starting from the parameters predicted from the neural work, we jointly optimize the geometry and appearance parameters with an approximated shading model in a high-performance differentiable framework. More specifically, we generate the fiber geometries procedurally with the geometric parameters (Sec.~\ref{sec:forward}) in Pytorch, where each fiber is represented with a quad facing to the camera ray. Then, the generated fibers are rasterized in a differentiable renderer, NVDiffRast, and shaded with our proposed shading model (Sec.~\ref{sec:DR}). Later, the loss is computed between the rendered and target images to drive the back-propagation. }
% \beibei{the image is so crowded. Make it the same feeling as Fig. 6.}}

%\added{If we use the approximated shading model, can we opt the geometry and the appearance together in Mitsuba3 with direct illumination?}
The parameters predicted by the network can produce rendered results roughly similar to the target image. However, there are still some geometry inaccuracies and color biases. Therefore, we jointly optimize the fiber geometric and appearance parameters to address those issues. However, due to the large parameter space and the long path of fibers, performing differentiable path tracing to optimize the geometry and the appearance parameters is infeasible. For this, our key insight is that an approximated shading model with parameters similar to the fiber shading model can be used for optimization, avoiding expensive path tracing. This approximated shading model can be used for final rendering applications with low time budgets or mapped to the original parameter space for high-quality rendering applications. We present our approximate differentiable fiber appearance model and then introduce a high-performance differentiable rendering framework for optimization.

\paragraph{Differentiable fiber appearance model}

To model the appearance of the fibers, we combine a single-scattering model and a multiple-scattering approximation. We use the single scattering fiber BSDF model by Chiang et al.~\shortcite{Chiang:2015:fur}, which is a near-field extension of Marschner's far-field hair BSDF~\cite{Marschner:2003:HairBCSDF} and is fairly standard in the computer graphics industry. Multiple scattering among the fibers is typically handled by path tracing (Monte Carlo random walks), but this remains too expensive and noisy for differentiable rendering purposes. To this end, we use a simple solution, approximating the multiple scattering with a diffuse term, which provides plausible quality and can be implemented in a differentiable form. {We do not use the Dual Scattering~\cite{Zinke:2008:dual}, because using Dual Scattering in fabrics struggles to fit multiple scatterings between fibers. Besides, its complex computation process makes it harder to pass the correct gradients compared to the simpler diffuse term, resulting in falling into local minima and higher performance costs.}

Our fabric shading model includes a BCSDF term and a diffuse term: $f(\win, \wout) =  f^{\mathrm{s}}(\win, \wout) + f^{\mathrm{d}}(\win, \wout)$. 
The fiber appearance parameters include the longitudinal roughness $\gamma_\mathrm{M}$, the azimuthal roughness $\gamma_\mathrm{N}$, and the albedo $C$ for the BCSDF term. 

The diffuse term is defined as a weighted sum of the Lambertian term under the original surface shading normal $\omega_n$ and the Lambertian term using the fiber normal $\omega_m$, similar to Jin et al.~\shortcite{Jin:2022:inverse} (yarn normal in their formulation):
\begin{equation}
     f^{\mathrm{d}}(\win, \wout) = w_\mathrm{d}\frac{k_\mathrm{d} \left \langle\win \cdot \omega_\mathrm{m} \right \rangle}{\pi \left \langle\win \cdot \omega_\mathrm{n} \right \rangle} + (1-w_\mathrm{d})\frac{k_\mathrm{d} }{\pi},
\end{equation}
here, $w_d$ denotes the blending weight, which is typically set to 0.5 in practice, and $k_\mathrm{d}$ represents the diffuse albedo.

\paragraph{High-performance differentiable rendering framework}

Based on the procedural geometric model and the approximated shading model, we build a high-performance differentiable rendering framework (see Fig.~\ref{fig:opt_pipleline}) for optimization. Our framework is based on NVDiffRast~\cite{Laine:2020:modular}, as it is a highly efficient and flexible differentiable framework. 

With the initialized geometric parameters from the network, we generate fibers procedurally with our geometric model in a differentiable manner (detailed in Sec.~\ref{sec:DRDetails}), and then rasterize the generated fibers in NVDiffRast with our appearance shading model, achieving differentiability across all three stages (generation, rasterization, shading). Because NVDiffRast is a rasterization-based framework, we replace the shadow ray casting with shadow maps (with the resolution set as $2\mathrm{K} \times 2\mathrm{K}$). In this way, we enable high-performance differentiable rendering of full fiber-level samples with these material parameters.

% \beibei{we do not say why we use these losses? Especially the first two. Is the RGB mean loss the same as the color loss in inverse SpongeCake?}

Next, the loss is computed, which drives the back-propagation through all three stages to optimize the desired parameters.  To measure the difference between the rendered image and the input image, we consider three components: a hierarchical VGG-19 Gram matrix loss $L_\mathrm{g}$, performed on downsampled images with $2\times2$, $4\times4$, and $8\times8$, an RGB mean color loss $L_\mathrm{c}$ and a prior loss $L_\mathrm{p}$.
Our final loss is defined as:
\begin{eqnarray}
    \label{eq:loss_opt}
    L_{\mathrm{opt}} &= &L_\mathrm{g} + L_\mathrm{c} + L_\mathrm{p},\\        
    L_\mathrm{g} &= &\bL_1(\mathrm{Gram}(I_\mathrm{down(2\times2)}), \mathrm{Gram}(R_\mathrm{down(2\times2)}))\nonumber \\
    &+ &\bL_1(\mathrm{Gram}(I_\mathrm{down(4\times4)}), \mathrm{Gram}(R_\mathrm{down(4\times4)}))\nonumber \\
    &+ &\bL_1(\mathrm{Gram}(I_\mathrm{down(8\times8)}), \mathrm{Gram}(R_\mathrm{down(8\times8)})),\\         
    L_\mathrm{c} &= &\bL_1(\mathrm{mean}(I),\mathrm{mean}(R)),\\
    L_\mathrm{p}&=&-\log\left(\exp\left(-\frac{(\gamma_M-\mu_{\gamma_M})^2}{2\sigma_{\gamma_M}^2}-\frac{(\gamma_{M_0}-\mu_{\gamma_{M_0}})^2}{2\sigma_{\gamma_{M_0}}^2}\right) \right).
    \label{eq:prior}
\end{eqnarray}
Where $\mu$ and $\sigma$ are the mean and the variance of the Gaussian prior. The optimization settings are detailed in Sec.~\ref{sec:DRDetails}. {We find that the Gram loss $L_g$ improves the similarity of the structure and color of the recovered result to the input, the color loss $L_c$ reduces the color bias, and the prior loss $L_p$ enhances optimization robustness.}

\paragraph{Discussion.}
Our approximate appearance model is not novel by itself; we use existing techniques with a few modifications. Our main contribution here is that this combination enables a differentiable framework to capture the geometry and the appearance of fabrics at the fiber level, which has not been possible before. More advanced techniques for either of them might further improve the quality, and we leave it for future work.

\subsection{Appearance refining with differentiable path tracing}
\label{sec:DPT}
%-----------------------------------%

With the joint geometry-appearance optimization, our method is able to produce plausible results. However, the estimated parameters are defined on an approximate appearance model -- a single scattering model together with a blended diffuse term. Unfortunately, this model can not be used directly in high-quality production rendering since a path tracing rendering with a single scattering hair BSDF can produce higher-fidelity results. Therefore, we propose to map the parameters of the approximate shading model to the parameters of the hair BSDF with a differentiable path tracing. Note that the similarity between the approximate appearance model and the target hair BSDF allows a good initialization of the parameters and makes the optimization more stable. 

Specifically, we map the single scattering albedo $C$, the diffuse albedo $k_\mathrm{d}$ and the roughnesses ($\gamma_M$, $\gamma_N$) produced by the differentiable optimization to the single scattering albedo $C$ and roughnesses ($\gamma_M$, $\gamma_N$), conditioned on all the other parameters. Note that we freeze all the fiber geometric parameters and refine the appearance parameters only in this step. 

We first initialize the single scattering hair BSDF parameters, setting $C^0 = \frac{\hat{C} + \hat{k}_\mathrm{d}}{2}$, $\gamma_M^0 = \hat{\gamma}_M$ and $\gamma_N^0 = \hat{\gamma}_N$. Then, we perform differentiable path tracing~\cite{Vicini:2021:PathReplayBackprop} on fibers with the single scattering model. Next, we compute the loss between the rendered and input images and use the loss to optimize the single scattering parameters. Here, we use the same loss (Eqn.~(\ref{eq:loss_opt}) to Eqn.~(\ref{eq:prior})) as the previous section. The optimization details are shown in Sec.~\ref{sec:DPTDetails}.

%% file: sec6_path.tex
\section{A framework for efficient fiber path tracing}
\label{sec:path}
%---------------------------------------------%

\myfigure{patch}{fig_two_scale_PT.pdf}{In the preprocessing step, we pre-generate the fiber geometries for a small patch and construct a hierarchy for the fiber geometries. The size of each patch is the same as a yarn pattern. For an input mesh, its surface is covered by repeating the pre-generated patch, {as shown in (a).} During rendering, the macro-scale position and the directions in the world space are transformed into the patch space. In the patch space (micro-scale), we find the exact intersection (green dot) by intersecting the ray with the fiber geometry. Then, we perform the path tracing in the patch space with the next event estimation until leaving the surface, as shown in (c).}

%We have presented an efficient differentiable fabric shading model based on a single scattering together with a diffuse term, which is suitable for optimization. However, it still produces lower quality than path tracing. To ensure the high quality of our data and bridge the domain gap between the synthetic and real data, we use path tracing for \beibei{training dataset rendering.} %final rendering of the resulting fabrics with the appearance model. 

For final fabric rendering, we need an efficient rendering solution. There are several challenges to render fabrics at the fiber level in human-scale scenes. First, the storage: given a fabric with size $100 \mathrm{cm} \times 100\mathrm{cm}$, there are (depending on parameters) about 20 million fibers with 2 billion vertices. Hence, we cannot realistically generate all fibers explicitly. Furthermore, we need to map our centerline curves onto the underlying object shape. To our knowledge, no existing off-the-shelf rendering method can achieve our goal; shell mapping \cite{Porumbescu:2005} could be considered, but its requirement of detailed discretization of the shell into tetrahedra is quite expensive. Therefore, we present a high-performance path-tracing framework implemented in Optix for fabric rendering at the fiber level. We introduce a patch-space fiber geometry generation method followed by our two-scale path tracing framework.  

\mycfigure{forward}{fig_forwardComp.pdf}{ Comparison between Irawan and Marschner~\shortcite{IrawanAndMarschner2012}, Jin et al.~\shortcite{Jin:2022:inverse} and our forward model on several fabrics for distant and close-up views. {We render our images by path tracing.} Our model produces results close to other models at the distant view {and shows fiber-level details at the close-up view. Note that we know the previous surface-based models cannot express fiber-level details. Here, we express that our model can also obtain the correct macro-scale appearance.} }
% \beibei{are we using path tracing? or are we using the approximate shading model. Here, we need to make it very clear, what's the purpose of this figure.}  }

\subsection{Patch-space fiber geometry generation}

To align the centerline curve with an arbitrary shape, we propose to define fiber geometry in \emph{patch space}. We treat each yarn pattern as a patch (tile) and cover the input shape by repeating these patches, considering the physical size, as shown in Fig.~\ref{fig:patch}. 

For each patch, given a set of fiber geometric parameters, we generate the fibers on a plane as shown in Fig.~\ref{fig:patch} (c). For each fiber, we use 40 vertices per yarn segment and then define a cylinder with two adjacent vertices. We also build a bounding volume hierarchy for these fibers, which will be used for ray-fiber intersection during rendering. Each patch only costs a  small amount of storage (0.74 MB + 1 MB for plain weave, 1.54 MB + 1.50 MB for twill, and 4.26 MB + 2.50 MB for satin).  The patches are instanced, so that an infinite grid of them supports ray intersections without having to store the patch copies.

% , which is simpler than the object space 

% \beibei{Note that the offset of yarns between the recovered image and the input is normal. This is due to a small discrepancy between the predicted yarn count and the input.}}

%Using the synthetic data as a target, our method can achieve recovery results that are similar to the target both microscopically and macroscopically.
\subsection{Two-scale path tracing}

Given the patch-space fiber geometry, we further propose a two-scale path tracing approach: a macro scale for the fabric surface, and a micro scale for the fibers. 

We shoot rays from the camera and perform the intersection with the macro surface, resulting in a location $x$ and a camera ray direction $\omega_\mathrm{o}$. Then, we sample a light ray with direction $\omega_\mathrm{i}$ by sampling the light sources. Next, we transform both the location $x$ and the directions $\omega_\mathrm{o}$ and $\omega_\mathrm{i}$ into patch space, and compute the intersection $\bar{x}$ with the pre-generated patch. The directions in the patch space are denoted as $\bar{\omega}_\mathrm{i}$ and $\bar{\omega}_\mathrm{o}$, respectively.

Next, we start path tracing at the micro-scale. We sample the hair BSDF~\cite{Chiang:2015:fur} to bounce the ray starting from $\bar{x}$ with direction $\bar{\omega}_\mathrm{o}$ and perform the next event estimation with the local light ray $\bar{\omega}_\mathrm{i}$ until the ray leaves the micro surface. We transform the ray into world space and continue the macro-path tracing for global illumination.

% We formulate the two-scale path tracing as follows

%{Similar to the idea of layered BSDF~\cite{Guo:2018:layered}},
We formulate the two-scale path tracing as follows:
% , where the micro-scale path tracing can be treated as a BSDF in the macro scale:
% \begin{eqnarray}
%     L_\mathrm{o}\left(x,\omega_\mathrm{o}\right) &=& \int L_\mathrm{i} \left(x,\omega_\mathrm{i}\right)
%     \rho \left(\omega_\mathrm{i},\omega_\mathrm{o}\right)
%     \mathrm{cos} \left(\omega_\mathrm{i} \right) \dd \omega_\mathrm{i}, \\
%     \rho \left(\omega_\mathrm{i},\omega_\mathrm{o}\right) &=& 
% \bar{L}_\mathrm{o}\left(\bar{x},\bar{\omega}_\mathrm{o},\bar{\omega}_\mathrm{i}\right), \\
% \bar{L}_\mathrm{o}\left(\bar{x},\bar{\omega}_\mathrm{o},\bar{\omega}_\mathrm{i}\right) &=& \int \bar{L}_\mathrm{i} \left(\bar{x},\bar{\omega}_\mathrm{i}\right)  
% f^s\left(\bar{\omega}_\mathrm{i}',\bar{\omega}_\mathrm{o}\right)
%     \mathrm{cos} \left(\bar{\omega}_\mathrm{i}' \right) \dd \bar{\omega}_\mathrm{i}',
% \end{eqnarray}
\added{
\begin{eqnarray}
    L_\mathrm{o}\left(x,\omega_\mathrm{o}\right) &=& \int 
\bar{L}_\mathrm{o}\left(\bar{x},\bar{\omega}_\mathrm{o},\bar{\omega}_\mathrm{i}\right)  \dd \omega_\mathrm{i}, \\
\bar{L}_\mathrm{o}\left(\bar{x},\bar{\omega}_\mathrm{o},\bar{\omega}_\mathrm{i}\right) &=& \int \bar{L}_\mathrm{i}\left(\bar{x},\bar{\omega}_\mathrm{i}\right)   
f^s\left(\bar{\omega}_\mathrm{i}',\bar{\omega}_\mathrm{o}\right)
    \mathrm{cos} \left(\bar{\omega}_\mathrm{i}' \right) \dd \bar{\omega}_\mathrm{i}',
\end{eqnarray}}
\added{where $f^s$ is the hair BSDF. Note that $\bar{L}_o\left(\bar{x},\bar{\omega}_\mathrm{o},\bar{\omega}_i\right)$ means the radiance of location $\bar{x}$ from direction $\bar{\omega}_\mathrm{o}$ with next event estimation (light direction set as $\bar{\omega}_\mathrm{o}$).}

We compare our rendered results with Jin et al.~\shortcite{Jin:2022:inverse}, Irawan and Marschner \shortcite{IrawanAndMarschner2012} on different patterns in Fig.~\ref{fig:forward}. {Our model matches their appearance in the distant view and additionally supports detailed results in the close-up view.} 

%% file: sec7_impl.tex
\section{Implementation details}
\label{sec:details}
%-----------------------------------%

%\subsection{Capture configurations}
%\label{sec:captureDetails}

%---------------------------------%
\subsection{Fabric network}
\label{sec:netDetails}
%---------------------------------%

\myfigure{patterns}{patterns.pdf}{Several fabric patterns studied in this paper.}

\paragraph{Dataset generation}
%Our procedural geometric model can generate a large amount of synthetic data mapping images to parameters. We use \added{the shading model of  \shortcite{Chiang:2015:fur}}, and render these images by path tracing. 

We currently support four weave patterns (twill0, twill1, satin, and plain weave, shown in Fig. \ref{fig:patterns}), together with 90-degree rotations of the satin and two types of twill, giving seven patterns in total. We generate 2,000 images for each pattern (14,000 images in total) by sampling the parameters of our geometric and appearance model, as shown in Table~\ref{tab:sample} . In the dataset, $90\%$ is used for training, and the remaining data is used for testing. We sample the parameter space with some priors from background knowledge. For example, compared to plain weave and twill, satin tends to have lower roughness, fiber twisting, yarn radius, and larger $u_{\mathrm{max}}$ and height field scaling factor. We render the training data using our CUDA-based path tracing framework (shown in Sec.~\ref{sec:path}), as path tracing can generate higher-quality renderings than rasterization. {We use 1024 spp, and the maximum bounce depth is 64. It takes a total of 268 hours to generate this dataset on a single NVIDIA RTX 3090 GPU.}
% \beibei{do we have a training and testing split? we have 1600 * 7 data in total? what's the resolution for rendered image? what's the bounce for path tracing and the spp? how long does it take? make sure this can be reproduced!!}

\paragraph{Training}
Our network is implemented in the PyTorch framework. We use the Adam solver, with a learning rate set to 0.0001. The training samples are fed into the network in a batch size of 8. Only fully connected weights are updated during training (VGG weights are frozen). Training takes five hours on a single NVIDIA A40 GPU.

% \paragraph{Estimate}
% We do not predict the yarn segment length directly but estimate the yarn counts $n^h, n^v$ instead. The yarn segment length can then be derived from the yarn count with $l_h = \frac{L_h}{n^v}$, where $n_h$ is the yarn count per yarn segment, determined by the yarn pattern.  

\subsection{Differentiable rasterization optimization}
\label{sec:DRDetails}

\paragraph{{Fiber geometry generation}}
Given the geometric parameters, we generate fibers around each centerline curve, where each fiber is represented with $10$ vertices per yarn segment. Between adjacent vertices, we generate a quad facing the camera direction. The normals at shading points are calculated by bilinear interpolation from the vertices. The generated mesh is rendered with NVDiffRast \cite{Laine:2020:modular}. We render our image at double resolution of the target image and filter down, to reduce the artifacts in the rasterization. 

\paragraph{Optimization settings}
During differentiable rasterization, we use the Adam optimizer with a learning rate of 0.01. To make the optimization more robust, we set a Gaussian prior loss on the longitudinal roughness $\gamma_{M}$ and the primary reflection roughness $\gamma_{M_0}$. The parameter settings for the prior loss are explained in Table~\ref{tab:prior}. We run for 75 iterations, which takes about ten minutes on an NVIDIA RTX 3090 GPU.

\subsection{Differentiable path tracing}
\label{sec:DPTDetails}

We perform the differentiable path tracing optimization in Mitsuba 3~\cite{jakob2022mitsuba3}, where the maximum depth for global illumination is set as 64 and the sample rate is set to 1024 for rendering and 256 for differentiable simulation. We use the Adam optimizer with a learning rate of 0.02. We run for 50 iterations, which takes about five minutes on an NVIDIA 3090 GPU.

%\section{Implementation details}

%\subsection{Chiang's fiber scattering model}
% We have the following precomputated tables: average forward and backward scattering attenuation $a_f$ and $a_b$, average forward and backward scattering variances $\beta_f$ and $\beta_b$, average backscattering attenuation $\Bar{A}_b$, average backscattering standard deviation $\sigma_b$, average three azimuthal functions over the front scattering hemisphere $N^G_R$,$N^G_{TT}$ and $N^G_{TRT}$. For each precomputation table, the resolution is set as 32. And the range of $C$,$\gamma_M$,$\gamma_N$ is $[0,1]$, the range of $\theta_d$ is $[-\frac{\pi}{2},\frac{\pi}{2}]$.
% \subsection{Geometry generation for optimization}
% After generation of control points by our geometric model. we generate two triangles facing the camera based on the fiber radius to form a camera facing quad and 10 quads make up one fiber. 

% For each vertex from quad, its tangent is obtained by subtracting two adjacent control points. And its normal is computed by moving upwards from the horizontal direction, which is obtained by taking the cross product of the difference between a control point and one of its adjacent vertices, and the tangent.We found that moving upward by 0.9 times the unit length from the horizontal direction results in interpolated normal that closely resemble the actual shape of the cylinder's fiber.
% (\added{Zixuan:We need a figure here to illustrate what we have done.})

%% file: sec8_results.tex
%-----------------------------------%
\section{Results}
\label{sec:results}
%-----------------------------------%
\mycfigure{syn_result}{syn_result.pdf}{Our recovery results on synthetic data. Each row displays the recovery result of a synthetic fabric after each step of our method, followed by the rendering of the draped cloth mesh at the end. Note that the result from joint geometry-appearance optimization via an approximated shading model shows a coarse and stiff look. However, after refining the appearance with differentiable path tracing, this effect is eliminated, resulting in a more translucent appearance. }

\mycfigure{real_result}{real_result.pdf}{Our recovery results on real data. Using real data as the input, our method can accurately predict and optimize parameters to match the input visually. Our joint geometry-appearance optimization using an approximate shading model produces an appearance that lacks translucency. Our further appearance refining improves this and results in an appearance closer to the real image. The rendered results on the draped cloth mesh also exhibit a plausible appearance. }
 %\myfigure{synthetic}{synthetic2.pdf}{Given synthetic input images, our neural network estimation can predict parameters that approach the appearance of the input. Using the optimization further improves the accuracy. Our results on the draped cloth mesh match the ground truth closely. }

%\myfigure{real}{real.pdf}{Given an input image captured with our measurement configuration, our inverse model is able to produce closely matching results. The rendered results on the draped cloth mesh also show a natural appearance.} 

%\myfigure{randomIntial}{randomInitial.pdf}{Comparison between a random initialization and our network outputs as an initialization.} 

% \added{fix the boring envmap.}
We first show the results of our procedural parameter estimation model on both synthetic and real inputs. Next, we show the impact of some critical components (neural network, joint optimization of geometric and appearance parameters with differentiable rasterization (DR), and refining of the fiber appearance parameters with differentiable path tracing (DPT)) in our inverse model. Lastly, we compare our inverse model with previous works~\cite{Jin:2022:inverse,Wu2019modeling}.
%More discussions and limitations are shown in the supplementary.

%general procedural method~\cite{Shi:2020:MATch} and a stationary SVBRDF recovery method~\cite{henzler2021generative}. 
%\revise{The capture of their inputs is shown in the supplementary material. 

 \subsection{Parameter estimation results}

%\added{compare with our inverseSpongeCake}
%distant view / close view

\input{table_syn_error}

\paragraph{Synthetic data.}
In Fig.~\ref{fig:syn_result}, we estimate the parameters on several kinds of synthetic fabrics with our method. Given the input images rendered with path tracing, our network recovers the overall geometry and appearance, but with some inaccuracy, such as color bias. Performing a joint geometry-appearance optimization with differentiable rasterization further enhances the accuracy of recovery. Despite using an approximate shading model, the rendered results of the estimated parameters become similar to the input. However, the rendered results cannot show the translucent appearance, due to the approximate multiple scattering. After mapping the parameters of the approximate shading model to the accurate fiber single scattering model together with path tracing for multiple scattering, the rendered results can produce closely matching results to the input images by high-quality rendering. We also show the rendered results on the draped cloth mesh and compare them with the ground truth. Our method can not only match the micro-scale appearance but also achieve a similar appearance on the meso-scale rendering compared to the ground truth. % because we constrain the parameter ranges and apply priors during optimization to enhance consistency between the recovered results and the ground truth at both the meso and micro scales.
The error between the recovered parameters and the ground truth is shown in Table~\ref{tab:error}, and the columns in the table from left to right correspond to each fabric from top to bottom in Fig.~\ref{fig:syn_result}. Most of the recovered parameters show very small differences from ground truth. In a few cases, certain parameters (such as $\gamma_N$ for Plain 1 and Twill 3, and $\gamma_\mathrm{M_0}$ for Twill 1) differ from the ground truth due to ambiguity in the recovery process. However, the rendered images are visually very similar. Note that we do not perform exposure optimization on our synthetic data because all our synthetic data uses the same known exposure. More results are shown in the appendix.
% \beibei{we may want to give a zoom-in of the DR. } \beibei{the cut in the last row is very obvious. still needs to be fixed. are they highlight? }
% \beibei{please let me know what's the rate of getting a satisfying results? please compress this figure, too large. }

\paragraph{Real data.}
In Fig.~\ref{fig:real_result}, we perform parameter estimation on measured data, including three typical patterns. Since there are no ground-truth parameters for measured data, we compare the visual match between the input image and the rendered image with the estimated parameters. The estimated results of the network (second column) match the real data in terms of pattern types, yarn counts, yarn radius, and fiber counts, but still have noticeable biases in both appearance and geometry. Our differentiable rasterization optimization reduces these biases. {Our differentiable path tracing optimization further refines the optical parameters, avoiding the coarse and stiff appearance of approximate models, resulting in a more translucent rendering.} More results are shown in the appendix. The renderings of the draped cloth mesh also show plausible appearances. 
% \beibei{we need to zoom in the DR one, as it's very important to show that the DPT is better.}
% \paragraph{\added{Discussion}}
% \added{In our two optimization steps, we use the image at the micro scale as the target, which may lead to discrepancies between the recovered parameters and the ground truth at the meso scale. To address this, we constrain the parameter ranges and apply priors during optimization to enhance consistency between the recovered results and the ground truth at both the meso and micro scales. When using synthetic data as the target, our recovered results closely match the ground truth at the mesoscale. When using real data as the target, the recovered results at the mesoscale are also reasonable.}

\subsection{Ablation study}

\myfigure{Ablation_shading_model}{Ablation_shading_model.pdf}{{Ablation study for the shading model. Initiated by the network, we demonstrate the recovery fabrics after the process of differentiable rasterization under identical settings. \textit{Single scattering only} cannot express the color change caused by multiple scattering. On the contrary, \textit{Diffuse only} loses highlights on the yarns. Combining the two can make the result match the input better.}}

\myfigure{Ablation_component}{ab_component.pdf}{Ablation study for the components in optimization. Using the same input capture, we demonstrate the impact of each component on the recovery of fabrics after network prediction. Our neural network estimation can predict both the geometry and appearance parameters of the input, but there is still bias in the geometry and appearance. Our differentiable rasterization optimization reduces the bias in geometry and appearance observed between the rendering result and the input (note the vertical
streak of highlight in the blue box). However, the appearance rendered after the differentiable rasterization optimization {relies on an approximate shading model, which produces a coarse and stiff look}. Following the differentiable path tracing optimization, the recovered fabrics can be rendered with more translucency that better fits the inputs. }
% \added{find other examples without DR.} }

\myfigure{Ablation_loss}{ab_loss.pdf}{{Ablation study for the Gram matrix loss and the color mean loss in optimization. Starting from the same settings, we present the recovery results of the real capture. The \textit{Gram matrix loss only} results in color bias, while using \textit{RGB mean loss only} is only able to approximately match the color but leads to incorrect geometry and appearance. Combining the two can better achieve results that match the inputs.}}

\myfigure{Ablation_prior}{ablation_prior.pdf}
{{Ablation study for the prior. Based on gram loss and color mean loss, we compared the impact of adding prior loss on the recovery results of real satin. The prior loss ensures that the parameters are optimized towards the correct range, resulting in a smoother and more realistic appearance of satin.}}

% \myfigure{failure_case}{failure_case.pdf}{\added{Failure case. Notice the appearance of truncation in the weft yarns.}}
% \emph{Impact of the geometric and shading models.}
% For geometric model: the hybrid way vs parabola function multiplied by a circular arc function only vs circular arc only. Showing the forward rendered results only in Fig.~\ref{fig:Ablation_geo}. \added{Zixuan: please work on this.}

\paragraph{The impact of the differentiable appearance model.} 
Our approximate appearance model plays a critical role in the joint optimization of the geometric and appearance fiber parameters. We validate its impact by comparing the recovered results with three different shading models: 1) \emph{single scattering + diffuse}, 2) \emph{single scattering only}, and 3) \emph{diffuse only}. {Note that we use the same settings (with network prediction and optimization settings) for these three cases for fairness, and present the recovery result after the network and differentiable rasterization.} 
% \beibei{are the results after three steps? or only the first two steps?}
%We employ the network for prediction in these three scenarios and then conduct differentiable rasterization optimization with these shading models under the same settings.
{As shown in Fig.~\ref{fig:Ablation_shading_model}, we use two captures of twill as the inputs. \emph{Single scattering only} fails to capture the color changes caused by multiple scattering, resulting in color bias.} Similarly, \emph{diffuse only} fails to produce any highlights on the yarns. {In contrast, our solution (\emph{single scattering + diffuse}) produces results that more closely resemble the inputs.} 
% \beibei{our ``solution'' can not match ``the input''. this is not true.}

\mycfigure{comparison_invsp}{comparison_invsp.pdf}{Comparison of our method with Jin et al.~\shortcite{Jin:2022:inverse} on synthetic red satin and synthetic yellow-green twill. Our method can match the inputs with less color bias at the meso scale. At the micro scale, our method can capture the exact yarn pattern more accurately and present fiber-level details.}
% \added{why red high lights of Jin et al.? From Milos.}}

%\emph{Impact of the three steps.}
%We validate the impacts of three main steps in our method: network prediction, optimization and the mapping network. 

%Optimization has a significant impact on the predicted results, as shown in Fig.~\ref{fig:ablation}.
%We compare the results with and without using FabricNet for initialization in Fig.~\ref{fig:randomIntial}. The parameters optimized with FabricNet match the inputs much better than optimization with random initialization, visually and quantitatively. After both loss curves converge, random initialization remains at a larger error than initialization with our FabricNet.

\paragraph{Impact of the inverse pipeline.}
 Our pipeline has three components: neural networks for parameter initialization, joint geometry-appearance optimization via an approximated shading model, and appearance refinement with differentiable path tracing. Each component in our pipeline significantly impacts the recovery results. We compare the results with and without each component in our optimization in Fig. ~\ref{fig:Ablation_component}. Using two captures of plain as inputs and starting from the network-predicted parameters, the rendered difference from the input image is noticeable. Our differentiable rasterization optimization further improves accuracy, although using an approximate shading model. With differentiable path tracing optimization, the recovery results are more translucent and therefore more realistic. If we skip the differentiable rasterization optimization and directly use the network-predicted parameters as the initialization for differentiable path tracing, a similar appearance can be recovered, but there is a noticeable geometric bias (note the vertical streak in the blue box). 
% \beibei{I am confused by this? how can we opt the geometry with the third step? or we do not opt the geometry?}

\paragraph{Impact of the loss in optimization.}
We further validate the influence of each loss in both joint geometry-appearance optimization and appearance refining, including the Gram matrix loss, the RGB mean loss, and the prior loss. To demonstrate its impact, we compare the recovered fabrics using different loss functions under identical settings. In Fig.~\ref{fig:Ablation_loss}, we use captures of one twill and one plain fabric sample as inputs. The \textit{Gram matrix loss only} introduces color bias, and using the \textit{RGB mean loss only} cannot adequately capture the geometry and appearance of the input. However, by combining the Gram matrix and the RGB mean losses, we achieve a better match with the input. {Including the Gram matrix loss improves the overall geometric structure and appearance. The RGB mean loss helps reduce the color bias. }To enhance optimization stability, we additionally include a prior loss. As depicted in Fig.~\ref{fig:Ablation_prior}, we {present} the recovery results for two real satin captures with and without the addition of prior loss on top of Gram matrix loss and color mean loss. The prior loss guides the optimization of parameters towards the correct range, ensuring a smooth appearance at both the micro and meso scales.

\subsection{Comparison with previous works}

\paragraph{Comparison with {a yarn-based model}.}
We compare our model with a lightweight yarn-level fabric recovery approach~\cite{Jin:2022:inverse} on synthetic data. We use path tracing with a single scattering model (BCSDF) with the same parameters to render images at two scales: a meso scale for Jin et al.~\shortcite{Jin:2022:inverse} shaped as a cylinder and a micro scale for our method as a plane. The light and other settings are set as their own configurations.

% \beibei{which figure?}
As shown in Fig.~\ref{fig:comparison_invsp}, we compare our method with Jin et al.~\shortcite{Jin:2022:inverse} on a twill and a satin. Both our method and theirs can produce plausible recovery results, but ours has less color bias. As expected, their model cannot capture detailed appearance in the zoomed-in view, whereas ours can reveal fiber-level details. \added{Note that at the micro scale, their method produces inaccuracies in appearance (as seen in the red highlights in the first row) and pattern (such as the twill in the second row). Since their input is at the meso scale and lacks precise yarn-level detail, the optimization process does not account for the correctness of appearance and pattern at the micro scale. }

%\added{can we still compare our model with Inverse sponge cake? since the setups are different.}
\myfigure{comp_Wu}{comp_withWu.pdf}{Comparison with another fiber based model \cite{Wu2019modeling}. Both previous work and ours achieve a plausible appearance. However, in the previous work, appearance parameters are manually tweaked, whereas ours are estimated automatically.}

\paragraph{Comparison with a fiber-based model.}
% \beibei{which figure?}
{We also compare our model with a {fiber-based model} \cite{Wu2019modeling} in Fig. \ref{fig:comp_Wu}. \beibei{They did not release code} and their method is non-trivial to reproduce due to its complexity. Additionally, the images they captured differ from our capture settings and have less clarity. Therefore, we chose one relatively clear twill image from their paper for comparison. Our method can achieve a plausible appearance similar to that of their result and has more fiber details. Note that their appearance model, based on a yarn-based model \cite{IrawanAndMarschner2012}, is not suitable for fiber rendering, while we use an accurate fiber model \cite{Chiang:2015:fur}. Furthermore, the appearance parameters they used are manually tweaked, while ours are estimated automatically.}

\myfigure{failure_case}{failure_case.pdf}{Failure case. The similar color between the weft and warp yarns raises the difficulties of recovery. }

\subsection{Discussion and limitations}

\paragraph{Fabric and pattern types.} Our model is currently designed for woven fabrics. We believe that it can also be applied to other types of fabrics (e.g., knitted fabrics) in the future. Moreover, we train our model on four typical patterns (seven with rotations). For other fabric patterns, the network needs to be trained on a larger dataset that includes the desired patterns, although no changes are needed for the optimization steps.

\paragraph{Flyaway fibers.} Our model follows previous procedural cloth models~\cite{Zhao:2016:Yarn} but we do not consider flyaway fibers. Adding flyaways can lead to a more realistic detailed appearance, though will make path tracing more costly. Flyaways also pose challenges in optimization. Adding flyaways is an important issue that needs to be addressed in the future.

\paragraph{Fabrics with \beibei{color} variations.} Fabrics could have color patterns or color variations, which is beyond the capability of our inverse model, although our model supports different weft and warp colors. Introducing spatially varying colors increases the parameter space significantly and makes the optimization more challenging. We leave it for future work. Nevertheless, spatially-varying effects can be applied to fabrics after the recovery. In Fig.~\ref{fig:teaser}, we further edit the appearance parameters using albedo maps to enhance the final appearance.

% \beibei{I merged the spatially varying and the blended.}
%Some fabrics in our capture are blended fabrics. However, we do not consider this blending because our model is procedural. In our model, the different colors in one yarn are estimate to one color which is similar to the blended of these colors. This result in an appearance similar to the captured blended fabrics images, and enabling procedural yarn models to support blending is an important future research topic.

%\added{\paragraph{Spatially-varying fabrics.}} \added{Spatially-varying effects are common in fabrics. However, because our model is procedural, it is difficult to recover these effects. Nonetheless, we can add spatially-varying effects to fabrics after the recovery. In Fig. \ref{fig:teaser}, we further edit the appearance parameters using albedo maps to enhance the final appearance.}

%\beibei{what do you mean here? what's the limitation here?} \added{\paragraph{Fiber-level noise.} Our fiber-level noise effectively simulates the scattering of fibers in real fabrics, while also to some extent replacing the effects of flyaway. However, due to downsampling our input images to 32x32 for better network parameter prediction, it becomes extremely challenging to predict the noise level from the network. So we initialize the initial value of the noise level at 1.0 for plain and twill, and 0.25 for satin, and correct it in the differentiable rasterization optimization.} 

\paragraph{Failure cases.} In our experiments, we noticed that our model might occasionally fail when the weft and the warp yarns have similar colors, which makes the boundary between yarns hard to see, as shown in Fig.~\ref{fig:failure_case}.

%% file: table_syn_error.tex
\begin{table*}
\centering
\caption{The error (in percentage) between the recovered parameters and the ground truth using synthetic data as the inputs. Each column in the table corresponds to a fabric listed from top to bottom in Fig.~\ref{fig:syn_result}. The recovery results of most parameters show very small differences from the ground truth.}
\begin{tblr}{
  column{even} = {c},
  column{1} = {c},
  column{3} = {c},
  column{5} = {c},
  column{7} = {c},
  column{9} = {c},
  hline{1,15} = {-}{0.08em},
  hline{2} = {-}{},
}
\textbf{Param.}     & \textbf{Plain 1} & \textbf{\textbf{Plain 2}} & \textbf{Plain 3} & \textbf{\textbf{Twill 1}} & \textbf{Twill 2} & \textbf{\textbf{Twill 3}} & \textbf{\textbf{\textbf{\textbf{Satin 1}}}} & \textbf{\textbf{\textbf{\textbf{\textbf{\textbf{\textbf{\textbf{Satin 2}}}}}}}} & \textbf{\textbf{\textbf{\textbf{\textbf{\textbf{\textbf{\textbf{Satin 3}}}}}}}} \\
$T$                 & 0\%               & 0\%                       & 0\%              & 0\%                       & 0\%              & 0\%                       & 0\%                                         & 0\%                                                                              & 0\%                                                                              \\
$C$                 & 0.66\%            & 6.58\%                     & 4.03\%            & 4.28\%                     & 9.13\%            & 3.84\%                     & 5.09\%                                       & 11.24\%                                                                          & 4.06\%                                                                           \\
$\gamma_M$          & 13.60\%           & 14.75\%                    & 5.70\%            & 11.85\%                    & 4.55\%            & 2.93\%                     & 0.05\%                                       & 0.23\%                                                                           & 0.41\%                                                                           \\
$\gamma_\mathrm{M0}$ & 8.05\%            & 5.92\%                     & 12.93\%           & 34.34\%                    & 7.73\%            & 15.57\%                    & 0.95\%                                       & 0.32\%                                                                           & 0.93\%                                                                           \\
$\gamma_N$          & 28.42\%           & 6.02\%                     & 19.19\%           & 5.92\%                     & 5.82\%            & 19.79\%                    & 14.64\%                                      & 12.81\%                                                                          & 3.35\%                                                                           \\
$\alpha$            & 3.21\%            & 5.63\%                     & 8.23\%            & 11.50\%                    & 9.57\%            & 3.71\%                     & 10.49\%                                      & 13.87\%                                                                           & 5.29\%                                                                           \\
$\beta$             & 4.00\%            & 3.08\%                     & 1.01\%            & 1.83\%                     & 1.17\%            & 5.54\%                     & 2.38\%                                       & 1.46\%                                                                           & 0.17\%                                                                           \\
$u_\mathrm{max}$    & 10.11\%           & 9.04\%                     & 27.61\%           & 2.43\%                     & 3.46\%            & 11.93\%                    & 19.25\%                                      & 14.00\%                                                                          & 5.61\%                                                                           \\
$e_\mathrm{yarn}$   & 0.60\%            & 0.92\%                     & 1.22\%            & 2.37\%                     & 2.43\%            & 6.03\%                     & 0.05\%                                       & 1.38\%                                                                           & 0.22\%                                                                           \\
$n$                 & 0.50\%            & 2.25\%                     & 1.00\%            & 0.75\%                     & 0.38\%            & 1.88\%                     & 2.13\%                                       & 10.63\%                                                                          & 2.25\%                                                                           \\
$m$                 & 1.17\%            & 2.22\%                     & 0.87\%            & 2.81\%                     & 4.09\%            & 2.74\%                     & 0.63\%                                       & 0.64\%                                                                           & 1.76\%                                                                           \\
$G$                 & 4.10\%            & 1.94\%                     & 2.50\%            & 6.81\%                     & 3.89\%            & 2.64\%                     & 3.68\%                                       & 1.81\%                                                                           & 0.90\%                                                                           \\
$Q$                 & 0.21\%            & 0.21\%                     & 0.27\%            & 0.08\%                     & 0.35\%            & 0.14\%                     & 0.20\%                                       & 0.07\%                                                                           & 0.14\%                                                                           
\end{tblr}
\label{tab:error}
\end{table*}

%% file: sec9_conclusion.tex
\section{Conclusion}
\label{sec:conclusion}
%-----------------------------------%

In this paper, we have presented a fiber-level procedural woven fabric parameter estimation approach from a single image captured with a low-cost microscope. For that, we propose a procedural geometric and shading model for fibers, allowing efficient differentiable rendering. Our work is the first to propose matching microscope photos using differentiable rendering with a micro-scale fiber geometric and appearance model, without manual tweaks. Our model is able to capture highly detailed fabrics, showing realistic appearances in both distant and close-up views. 

In the future, we plan to enhance the realism of our geometric model by including flyaways and extend the capabilities of our model for more effects, such as knitted and multi-ply fabrics. We also believe applying an extension of our proposed method to hair or fur reconstruction might be worthwhile.

%In this paper, we have presented a forward fabric shading model and an inverse procedural framework for estimating the model parameters, including a neural network and an optimization based on differentiable rendering. Our forward model can generate high-quality renderings at the yarn level, while remaining simple to implement, supporting efficient differentiable rendering and synthetic data generation. Our inverse framework can estimate fabric parameters that match the ground truth for the synthetic data, and remain close to the input images for measured data.

%We believe that our inverse framework will act as a foundation for future practical and accurate fabric capture methods. In the future, we plan to extend the capabilities of our forward and inverse models to more types of yarns, wider range of patterns, and more advanced effects such as transmission and knitted fabrics.

%\begin{acks}
%We thank the reviewers for their valuable suggestions. This work has been partially supported by the National Natural Science Foundation of China under grant No. 62172220. Ling-Qi Yan is supported by gift funds from Adobe, Dimension 5 and XVerse.
%\end{acks}

%% file: paper_supplementary.tex
%%
%% This is file `sample-authordraft.tex',
%% generated with the docstrip utility.
%%
%% The original source files were:
%%
%% samples.dtx  (with options: `authordraft')
%%
%% IMPORTANT NOTICE:
%%
%% For the copyright see the source file.
%%
%% Any modified versions of this file must be renamed
%% with new filenames distinct from sample-authordraft.tex.
%%
%% For distribution of the original source see the terms
%% for copying and modification in the file samples.dtx.
%%
%% This generated file may be distributed as long as the
%% original source files, as listed above, are part of the
%% same distribution. (The sources need not necessarily be
%% in the same archive or directory.)
%%
%% The first command in your LaTeX source must be the \documentclass command.

%timestamp

\definecolor{mygray}{rgb}{0.9,0.9,0.9}

\section{IMPLEMENTATION DETAILS}

\subsection{Prior loss}
To ensure the stability of joint geometry-appearance optimization and appearance refining, we set a Gaussian prior loss on the longitudinal
roughness $\gamma_{M}$ and the primary reflection roughness $\gamma_{M0}$:
\begin{equation}
    L_\mathrm{p}\quad=\quad-\log\left(\exp\left(-\frac{(\gamma_{M}-\mu_{\gamma_M})^{2}}{2\sigma_{\gamma_M}^{2}}-\frac{(\gamma_{M_{0}}-\mu_{\gamma_{M_{0}}})^{2}}{2\sigma_{\gamma_{M_{0}}}^{2}}\right)\right).
\end{equation}
Where $\mu$ and $\sigma$ are the mean and the variance of the Gaussian prior. For different pattern types, the respective $\mu$ and $\sigma$ values are shown in Table~\ref{tab:prior}.

\subsection{Parameters sampling}
We show the sampling distribution of our parameters space in Table~\ref{tab:sample}. Note that each type of fabric has only one fabric pattern $W$, while the other parameters have separate versions for weft and warp.

\section{MORE RESULTS}
\subsection{Synthetic data}
In Fig.~\ref{fig:syn_result_sup}, we show the results of our inverse model on synthetic data. Our inverse model, including the predict network, joint geometry-appearance optimization, and appearance refining, produces results that closely match the inputs. In meso-scale, our renderings with the draped cloth also match the ground truth.

\subsection{Real data}
In Fig.~\ref{fig:real_result_sup}, we provide more real data than in the main paper. All these images are captured by our measured configuration. Our method produces the recovery results that visually match the inputs. The renderings with the draped cloth mesh also show a plausible appearance.
\input{table_prior}
\input{table_sample}

\begin{figure*}[tb]
    \centering
    \begin{subfigure}[tb]{\linewidth}
        \centering
        \includegraphics[clip, width=\linewidth]{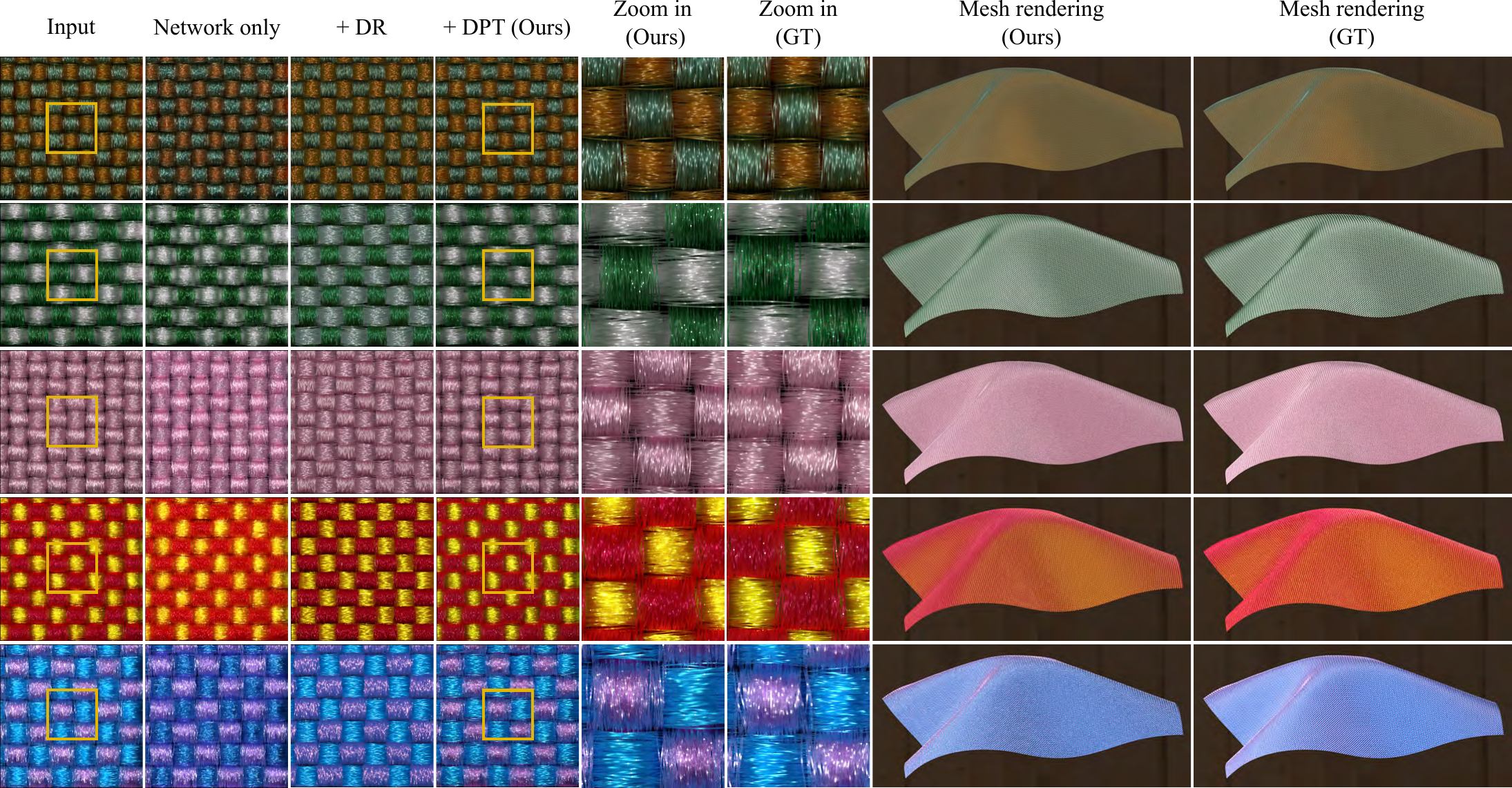}
        \caption{Plain}
        \label{fig:syn_sup_plain}
    \end{subfigure}
    \vskip 3mm
    \begin{subfigure}[tb]{\linewidth}
        \centering
        \includegraphics[clip, width=\linewidth]{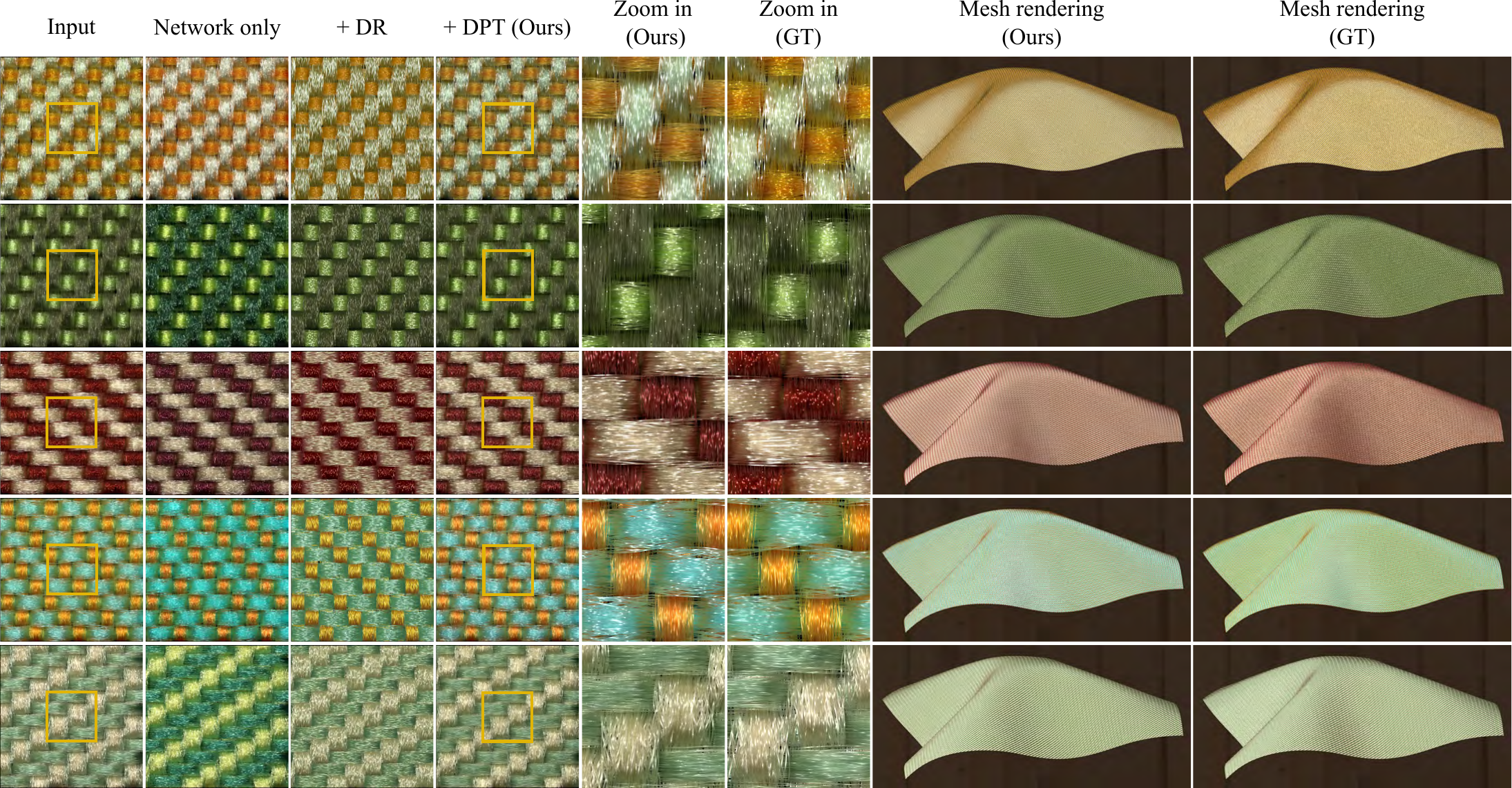}
        \caption{Twill}
        \label{fig:syn_sup_twill}
    \end{subfigure}   
\end{figure*}
\begin{figure*}[tb]\ContinuedFloat
\begin{subfigure}[tb]{\linewidth}
        \centering
        \includegraphics[clip, width=\linewidth]{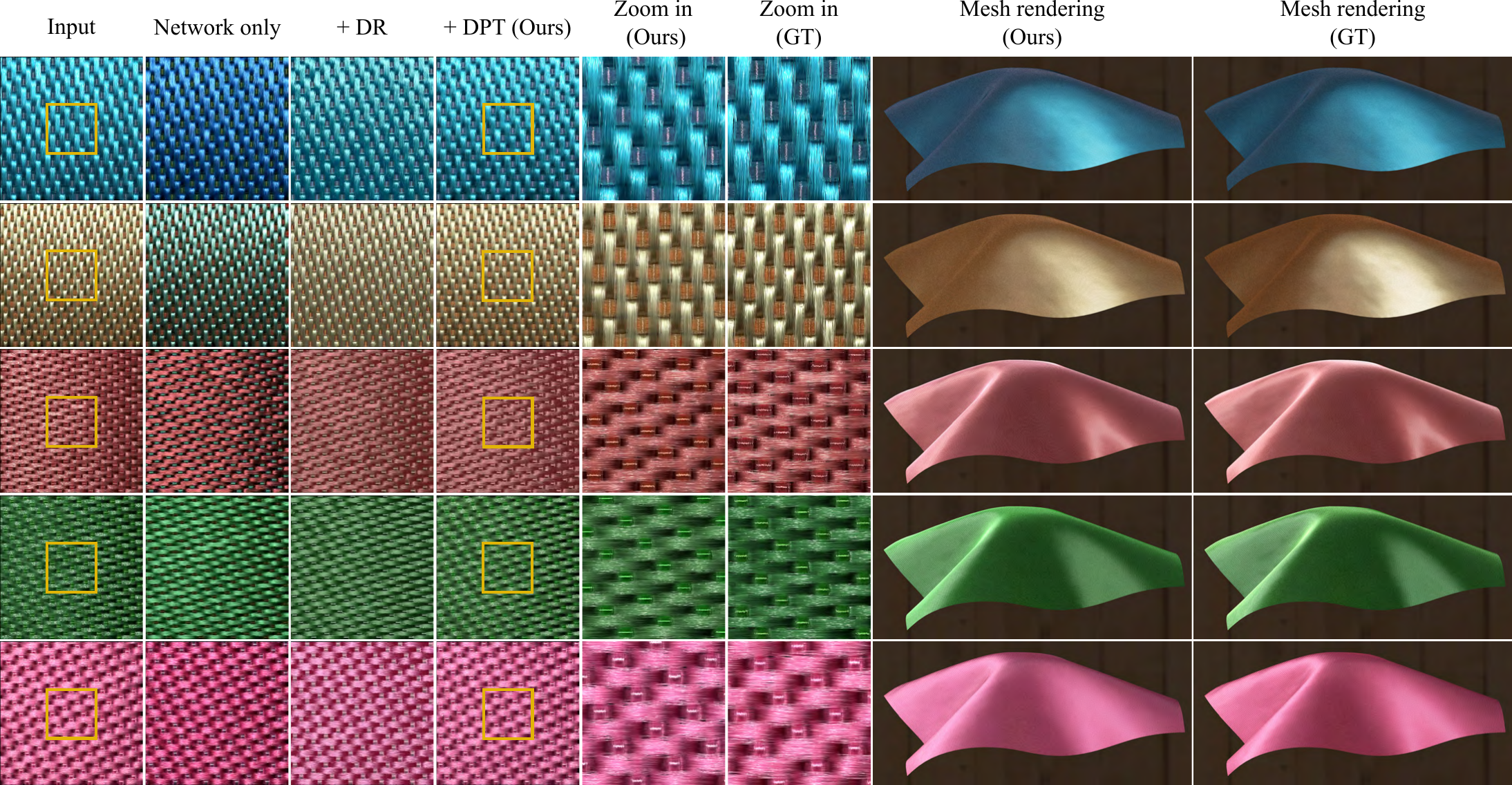}
        \caption{Satin}
        \label{fig:syn_sup_satin}
    \end{subfigure}
    \caption{Our recovery results on synthetic data. Each row shows the recovery result of a synthetic fabric after each step of our method, with the renderings on the draped mesh at the end.}
    \label{fig:syn_result_sup}
    \end{figure*}

\begin{figure*}[tb]
    \centering
    \begin{subfigure}[tb]{\linewidth}
        \centering
        \includegraphics[clip, width=\linewidth]{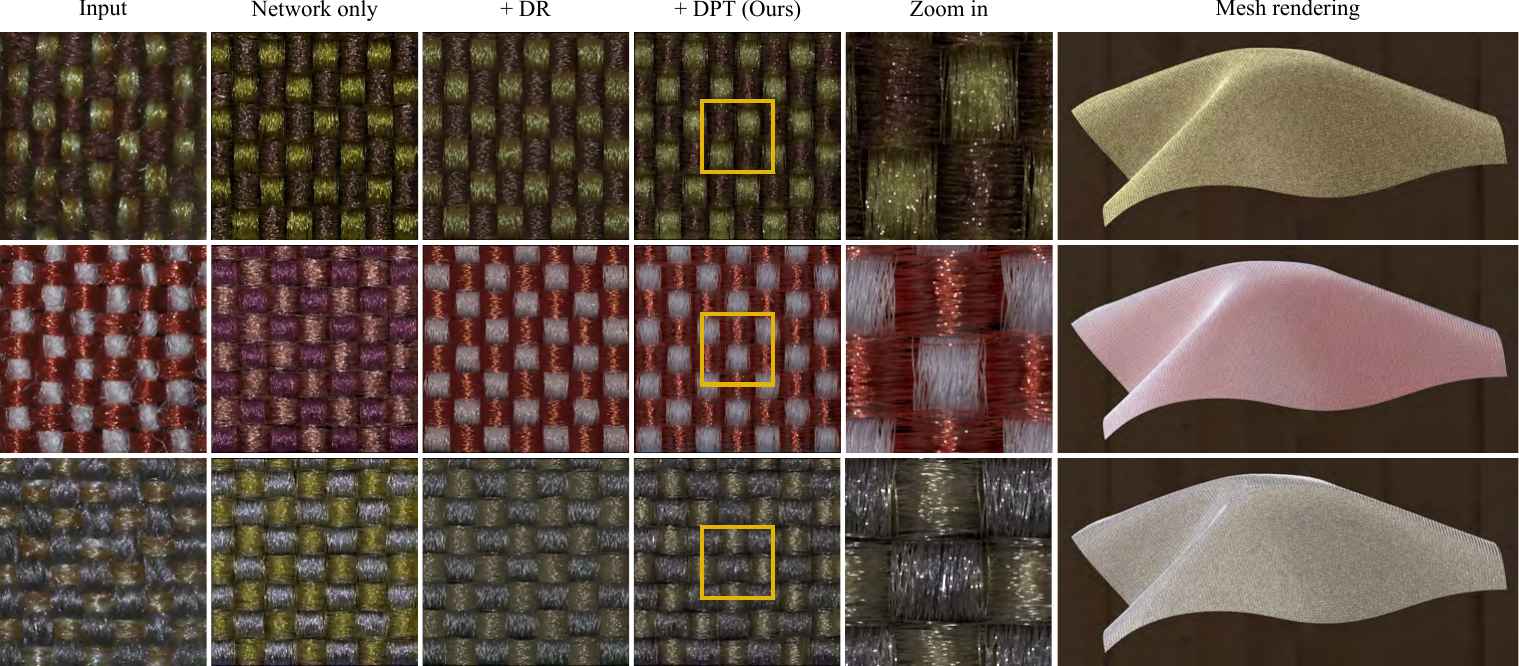}
        \caption{Plain}
        \label{fig:real_sup_plain}
    \end{subfigure}
    
\end{figure*}
\begin{figure*}[tb]\ContinuedFloat
\begin{subfigure}[tb]{\linewidth}
        \centering
        \includegraphics[clip, width=\linewidth]{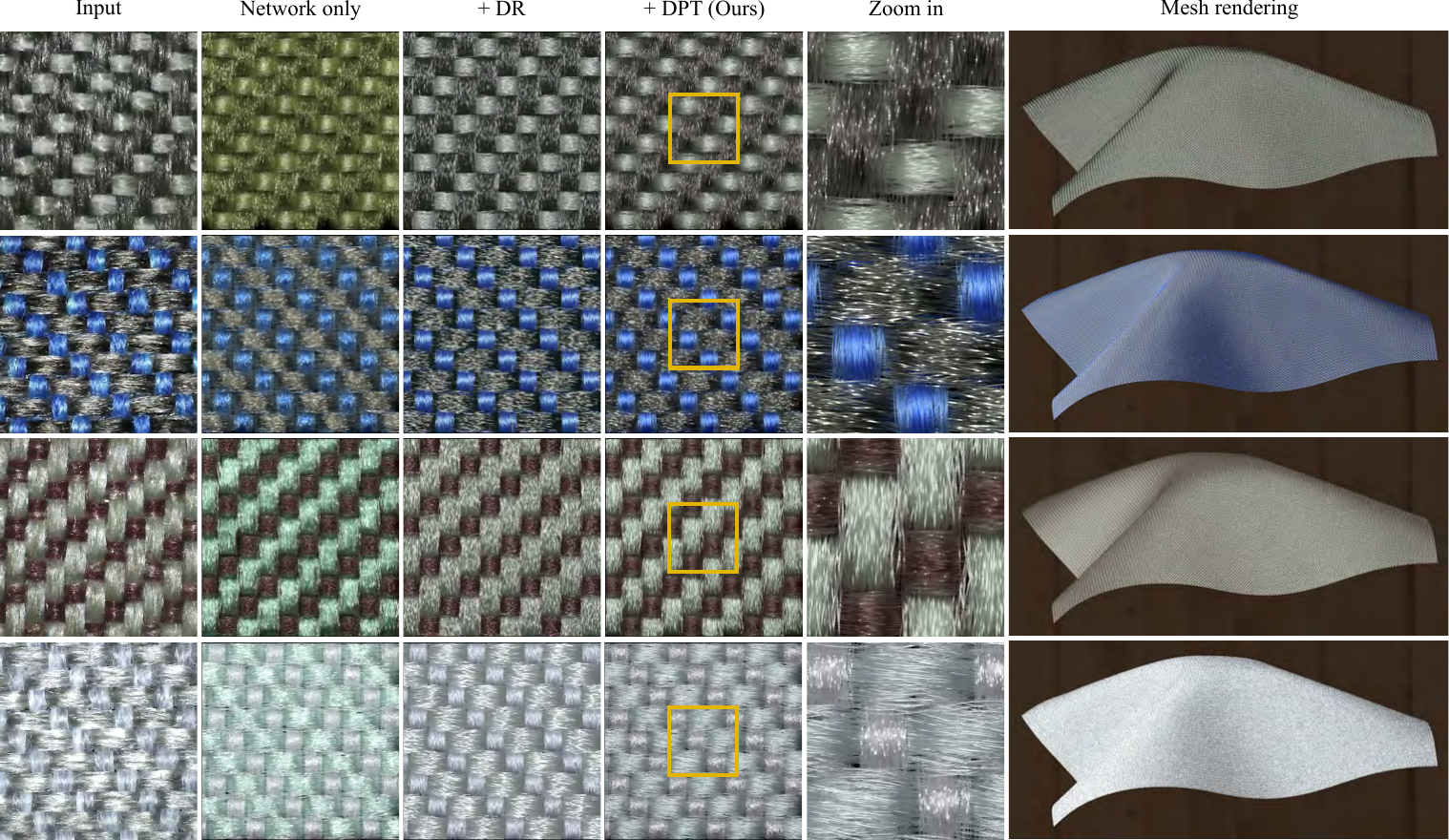}
        \caption{Twill}
        \label{fig:real_sup_twill}
    \end{subfigure}
    \vskip 3mm
\begin{subfigure}[tb]{\linewidth}
        \centering
        \includegraphics[clip, width=\linewidth]{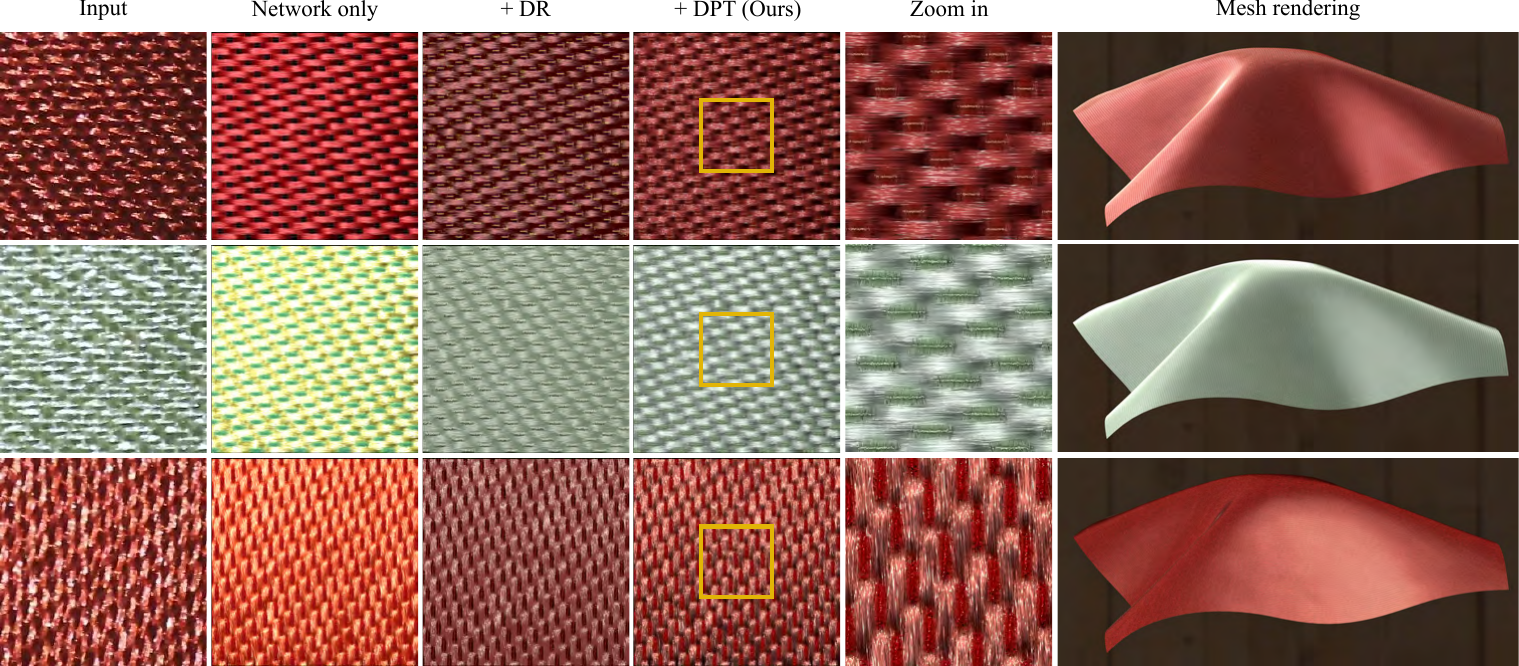}
        \caption{Satin}
        \label{fig:real_sup_satin}
    \end{subfigure}
    \caption{Our recovery results on real data. Using the measured images as the inputs, our inverse model produces recovery results that visually match the inputs. The renderings on the draped mesh also show plausible appearances.}
    \label{fig:real_result_sup}
    \end{figure*}
    
% \mycfigureTwoPages{syn_result_sup}{syn_result_sup_1.pdf}{syn_result_sup_2.pdf}{Our recovery results on synthetic data. Each row shows the recovery result of a synthetic fabric after each step of our method, with the renderings on the draped mesh at the end.}

% \mycfigureTwoPages{real_result_sup}{real_result_sup_1.pdf}{real_result_sup_2.pdf}{Our recovery results on real data. Use the measured images as the inputs, our inverse model produces recovery results that visually match the inputs. The renderings on the draped mesh also show plausible appearance.}
% \mycfigure{syn_result_sup}{syn_result_sup.pdf}{\revise{Our model can handle fabrics with different warp and weft colors.}}

% \mycfigure{real_result_sup}{real_result_sup.pdf}{\revise{Our model can handle fabrics with different warp and weft colors.}}

% \includepdf[pages=-,fitpaper=true]{syn_result_sup.pdf}

% test

% \input{sec_results_suppl}
%\input{sec_impl}

%%
%% The acknowledgments section is defined using the "acks" environment
%% (and NOT an unnumbered section). This ensures the proper
%% identification of the section in the article metadata, and the
%% consistent spelling of the heading.
% \begin{acks}
% To Robert, for the bagels and explaining CMYK and color spaces.
% \end{acks}

%%
%% The next two lines define the bibliography style to be used, and
%% the bibliography file.
%\newpage 
% \bibliography{paper}
% Appendix
%\appendix
%\input{paper_appendix}

% \endinput
%%
%% End of file `sample-authordraft.tex'.

%% file: table_prior.tex
\begin{table}
\centering
\setlength{\tabcolsep}{4pt}
\caption{The value of mean $\mu$ and variance $\sigma$ in our Gaussian prior loss. Note that since $\gamma_{M}$ and $\gamma_{M0}$ are defined within (0,1), the prior's effect on optimization for plain and twill is negligible. }
\begin{tabular}{ccccc}
% \hline
\toprule
pattern & $\mu$ ($\gamma_{M}$) & $\sigma$ ($\gamma_{M}$) & $\mu$ ($\gamma_{M_0}$) & $\sigma$ ($\gamma_{M_0}$)\\
% \hline
\midrule
Plain & 0.5 & 5 & 0.5 & 5 \\
% \hline
Twill & 0.5 & 5 & 0.5 & 5 \\
% \hline
Satin & 0.1 & 0.15 & 0.02 & 0.15 \\
% \hline
\bottomrule
\end{tabular}
\label{tab:prior}
\end{table}

%% file: table_sample.tex
\begin{table*}
    \centering
    \setlength{\tabcolsep}{4pt}
    \caption{Distributions used to sample the parameter space of our model. Except for the fabric pattern in the second row, all other parameters have separate versions for weft and warp. $\mathcal{V}(X)$ is a discrete uniform random variable on a finite set $X$. $\mathcal{N}(\mu, \sigma)$ represents a Gaussian distribution with mean $\mu$ and standard deviation $\sigma$. $\mathcal{U}(a,b)$ is a uniform distribution over the interval $(a,b)$.}
    \scalebox{0.88}{
    \begin{tabular}{ccc} 
    \toprule
        Parameter & weft Sampling Function & warp Sampling Function\\
    \midrule
        fabric pattern & \multicolumn{2}{c}{$\mathrm{W} = \mathcal{V}(\{0, 1, 2, 3, 4, 5, 6, 7\})$} \\
    \midrule
    \rowcolor{mygray} \multicolumn{3}{c}{Plain} \\
    \midrule
        fiber twisting & $\alpha = \mathcal{N}(0.0, 0.03)$ & $\alpha = \mathcal{N}(0.0, 0.03)$ \\
        yarn radius & $e_\mathrm{yarn} = \mathcal{U}(0.0525, 0.09625)$ & $e_\mathrm{yarn} = \mathcal{U}(0.0525, 0.09625)$\\
        yarn fiber count & $m = \mathcal{N}(200, 5)$ & $m = \mathcal{N}(200, 5)$ \\
        yarn count & $n = \mathcal{U}(6, 10)$ & $n = \mathcal{U}(5, 7)$\\
        fiber migration range & $G = \mathcal{N}(0.75, 0.03)$ & $G = \mathcal{N}(0.75, 0.03)$ \\
        fiber migration scale & $s = \mathrm{max}\{\mathcal{N}(0.2, 0.03), 0.01\}$  & $s = \mathrm{max}\{\mathcal{N}(0.2, 0.03), 0.01\}$ \\
        maximum inclination angle & $u_{\mathrm{max}} = \mathcal{U}(0.1, 1.5)$   & $u_{\mathrm{max}} = \mathcal{U}(0.1, 1.5)$\\
        heightfield scaling factor & $\beta = \mathcal{U}(0.3, 1.5)$  & $\beta = \mathcal{U}(0.3, 1.5)$ \\
        fiber albedo coefficients & $C = \mathcal{U}(0, 1)$  & $C = \mathcal{U}(0, 1)$\\
        longitudinal roughness &  $\gamma_{M} = \mathrm{max}\{\mathcal{N}(0.5, 0.1), 0.01\}$  & $\gamma_{M} = \mathrm{max}\{\mathcal{N}(0.5, 0.1), 0.01\}$\\      
        azimuthal roughness & $\gamma_{N} = \mathrm{max}\{\mathcal{N}(0.05, 0.01), 0.01\}$   & $\gamma_{N} = \mathrm{max}\{\mathcal{N}(0.05, 0.01), 0.01\}$\\
        primary reflection roughness & $\gamma_{M0} =\mathcal{U}(0.1, 0.15)$  & $\gamma_{M0} = \mathcal{U}(0.1, 0.15)$ \\ 
    \midrule
    \rowcolor{mygray} \multicolumn{3}{c}{Twill0} \\
    \midrule
        fiber twisting & $\alpha = \mathcal{N}(0.0, 0.03)$ & $\alpha = \mathcal{N}(0.0, 0.03)$ \\
        yarn radius & $e_\mathrm{yarn} = \mathcal{U}(0.0475, 0.108)$ & $e_\mathrm{yarn} = \mathcal{U}(0.0475, 0.0672)$\\
        yarn fiber count & $m = \mathcal{N}(200, 5)$ & $m = \mathcal{N}(200, 5)$ \\
        yarn count & $n = \mathcal{U}(9, 11)$ & $n = \mathcal{U}(6, 11)$\\
        fiber migration range & $G = \mathcal{N}(0.75, 0.03)$ & $G = \mathcal{N}(0.75, 0.03)$ \\
        fiber migration scale & $s = \mathrm{max}\{\mathcal{N}(0.2, 0.03), 0.01\}$  & $s = \mathrm{max}\{\mathcal{N}(0.2, 0.03), 0.01\}$ \\
        maximum inclination angle & $u_{\mathrm{max}} = \mathcal{U}(0.1, 1.5)$   & $u_{\mathrm{max}} = \mathcal{U}(0.1, 1.5)$\\
        heightfield scaling factor  & $\beta = \mathcal{U}(0.3, 1.5)$  & $\beta = \mathcal{U}(0.3, 1.5)$ \\
        fiber albedo coefficients & $C = \mathcal{U}(0, 1)$  & $C = \mathcal{U}(0, 1)$\\
        longitudinal roughness & $\gamma_{M} = \mathrm{max}\{\mathcal{N}(0.5, 0.1), 0.01\}$  & $\gamma_{M} = \mathrm{max}\{\mathcal{N}(0.5, 0.1), 0.01\}$\\    
        azimuthal roughness & $\gamma_{N} = \mathrm{max}\{\mathcal{N}(0.05, 0.01), 0.01\}$   & $\gamma_{N} = \mathrm{max}\{\mathcal{N}(0.05, 0.01), 0.01\}$\\  
        primary reflection roughness & $\gamma_{M0} =\mathcal{U}(0.1, 0.15)$  & $\gamma_{M0} = \mathcal{U}(0.1, 0.15)$ \\
    \midrule
    \rowcolor{mygray} \multicolumn{3}{c}{Twill1} \\
    \midrule
        fiber twisting & $\alpha = \mathcal{N}(0.0, 0.03)$ & $\alpha = \mathcal{N}(0.0, 0.03)$ \\
        yarn radius & $e_\mathrm{yarn} = \mathcal{U}(0.038, 0.055)$ & $e_\mathrm{yarn} = \mathcal{U}(0.045, 0.0672)$\\
        yarn fiber count & $m = \mathcal{N}(200, 5)$ & $m = \mathcal{N}(200, 5)$ \\
        yarn count & $n = \mathcal{U}(11, 13)$ & $n = \mathcal{U}(9, 11)$\\
        fiber migration range & $G = \mathcal{N}(0.75, 0.03)$ & $G = \mathcal{N}(0.75, 0.03)$ \\
        fiber migration scale & $s = \mathrm{max}\{\mathcal{N}(0.2, 0.03), 0.01\}$  & $s = \mathrm{max}\{\mathcal{N}(0.2, 0.03), 0.01\}$ \\
        maximum inclination angle & $u_{\mathrm{max}} = \mathcal{U}(0.1, 1.5)$   & $u_{\mathrm{max}} = \mathcal{U}(0.1, 1.5)$\\
        heightfield scaling factor  & $\beta = \mathcal{U}(0.3, 1.5)$  & $\beta = \mathcal{U}(0.3, 1.5)$ \\
        fiber albedo coefficients & $C = \mathcal{U}(0, 1)$  & $C = \mathcal{U}(0, 1)$\\
        longitudinal azimuthal roughness & $\gamma_{M} = \mathrm{max}\{\mathcal{N}(0.5, 0.1), 0.01\}$  & $\gamma_{M} = \mathrm{max}\{\mathcal{N}(0.5, 0.1), 0.01\}$\\     
        azimuthal roughness & $\gamma_{N} = \mathrm{max}\{\mathcal{N}(0.15, 0.01), 0.01\}$   & $\gamma_{N} = \mathrm{max}\{\mathcal{N}(0.15, 0.01), 0.01\}$\\ 
        primary reflection roughness & $\gamma_{M0} =\mathcal{U}(0.1, 0.15)$  & $\gamma_{M0} = \mathcal{U}(0.1, 0.15)$ \\
    \midrule
    \rowcolor{mygray} \multicolumn{3}{c}{Satin} \\
    \midrule
        fiber twisting & $\alpha = \mathcal{N}(0.0, 0.01)$ & $\alpha = \mathcal{N}(0.0, 0.01)$ \\
        yarn radius & $e_\mathrm{yarn} = \mathcal{U}(0.0053, 0.0143)$ & $e_\mathrm{yarn} = \mathcal{U}(0.0234, 0.0357)$\\
        yarn fiber count & $m = \mathcal{N}(200, 5)$ & $m = \mathcal{N}(200, 5)$ \\
        yarn count & $n = \mathcal{U}(11, 13)$ & $n = \mathcal{U}(9, 11)$\\
        fiber migration range & $G = \mathcal{N}(0.75, 0.03)$ & $G = \mathcal{N}(0.75, 0.03)$ \\
        fiber migration scale & $s = \mathrm{max}\{\mathcal{N}(0.2, 0.03), 0.01\} \cdot 0.8 $  & $s = \mathrm{max}\{\mathcal{N}(0.2, 0.03), 0.01\} \cdot 0.8 $ \\
        maximum inclination angle & $u_{\mathrm{max}} = \mathcal{U}(0.1, 1.5)$   & $u_{\mathrm{max}} = \mathcal{U}(0.1, 1.5)$\\
        heightfield scaling factor  & $\beta = \mathcal{U}(0.5, 1.5) $  & $\beta = \mathcal{U}(0.1, 0.5)$\\
        fiber albedo coefficients & $C = \mathcal{U}(0, 1)$  & $C = \mathcal{U}(0, 1)$\\
        longitudinal roughness & $\gamma_{M} = \mathrm{max}\{\mathcal{N}(0.4, 0.1), 0.01\}$  & $\gamma_{M} = \mathrm{max}\{\mathcal{N}(0.4, 0.1), 0.01\}$\\     
        azimuthal roughness & $\gamma_{N} = \mathrm{max}\{\mathcal{N}(0.05, 0.01), 0.01\}$   & $\gamma_{N} = \mathrm{max}\{\mathcal{N}(0.05, 0.01), 0.01\}$\\ 
        primary reflection roughness & $\gamma_{M0} =\mathcal{U}(0.1, 0.15)$  & $\gamma_{M0} = \mathcal{U}(0.1, 0.15)$ \\
    \bottomrule
    \end{tabular}}
    \label{tab:sample}
\end{table*}